\documentclass[%
 reprint,
 amsmath,amssymb,
 aps,
floatfix,
]{revtex4-2}

\usepackage[title]{appendix}
\usepackage{amsmath}   
 \usepackage{amsfonts}
\usepackage{amssymb}
\usepackage{multirow}
\usepackage{booktabs}
\usepackage{array}
\usepackage{graphicx}
\usepackage{float}
\usepackage[export]{adjustbox} 
\usepackage{subcaption}
\usepackage{subfiles}
\usepackage{braket}
\usepackage{svg}
\usepackage{placeins} 
\usepackage{tikz}
    \usetikzlibrary{quantikz}
\usepackage{dcolumn}
\usepackage{bm}
\usepackage{algorithm}
\usepackage{algpseudocode}
\usepackage{dcolumn}
\usepackage{bm}

\begin{document}

\preprint{APS/123-QED}

\title{Quantum machine learning for the quantum lattice Boltzmann method: Trainability of variational quantum circuits for the nonlinear collision operator across multiple time steps}

\author{Antonio David Bastida Zamora}
\altaffiliation{Also at Aix-Marseille University, France}
\affiliation{Quanscient Oy, Finland}

\author{Ljubomir Budinski}
\altaffiliation{Also at Faculty of Technical Sciences, University of Novi Sad, Serbia}
\affiliation{Quanscient Oy, Finland}

\author{Pierre Sagaut}
\affiliation{Aix Marseille Univ, Centrale Med, CNRS, M2P2 Laboratory, 13013 Marseille, France}

\author{Valtteri Lahtinen}
\altaffiliation{Also at School of Engineering Science, Lappeenranta–Lahti University of Technology, Finland}
\affiliation{Quanscient Oy, Finland}

\date{\today}

\begin{abstract}
This study investigates the application of quantum machine learning (QML) to approximate the nonlinear component of the collision operator within the quantum lattice Boltzmann method (QLBM). To achieve this, we train a variational quantum circuit (VQC) to construct an operator $U$. When applied to the post-linear-collision quantum state $\ket{\Psi_i}$, this operator yields a final state $\ket{\Psi_f} = U\ket{\Psi_i}$ that successfully replicates the nonlinear collision dynamics derived from the Bhatnagar-Gross-Krook (BGK) approximation. Within this framework, we present two distinct architectures: the R1 model and the R2 model. The R1 model is designed for quantum simulations that involve multiple time steps without intermediate measurements, focusing on accurately capturing nonlinear dynamics in continuous evolution. In contrast, the R2 model is tailored to achieve the high-precision reconstruction of the nonlinear operator for a single time step with an unitary operator.
\end{abstract}

\maketitle

\section{Introduction}
Researchers have long attempted to solve classical problems using quantum computers. While quantum computing seems ideal for applications such as chemistry and quantum many-body interactions, its applicability to macroscopic classical systems has also attracted significant interest. One of the broadest and most general problems being faced is the solution of partial differential equations on quantum computers for efficient and large-scale simulations. To solve this challenge, academic and industrial groups around the world have presented numerous quantum algorithm proposals: Linear solvers such as Harrow–Hassidim–Lloyd algorithm \cite{PhysRevLett.103.150502}, physics-informed quantum neural networks (QNN) \cite{panichi2025qpinns,berger2025teqpin}, imaginary time evolution \cite{kumar2024qitepde}, lattice Boltzmann method \cite{Budinski_2021}, lattice gas automata \cite{ZAMORA2025106476}, hybrids between lattice gas and lattice Boltzmann \cite{5bby-34zx}, finite element method \cite{alkadri2025quFEM}, Schrödingerization of partial differential equations \cite{Sasaki2025CLS}, quantum smoothed particle hydrodynamics \cite{au-yeung2025qSPH}, quantum annealing \cite{criado2023Qade,KUYA2024106238} and many others. Despite the numerous methods and inspiring combinations of some of them by various authors, a low-depth quantum algorithm that captures nonlinear terms while remaining coherent without measurements has yet to be discovered. 

One of the most promising algorithms mentioned above is the lattice Boltzmann method (LBM). This computational method was developed to address technical problems encountered in lattice gas automata (LGA), a promising Boolean-based algorithm developed in the 1970s and 1980s. Despite LGA's success in modelling nonlinear partial differential equations, such as the Navier-Stokes equation, using simple collision rules and propagation steps, the method was found to include non-physical behaviour and require a large number of executions to smooth the results. Since its creation, LBM has been successfully applied to various fields, including electromagnetics \cite{PhysRevE.96.063306}, computational fluid dynamics \cite{qin2023simplifiedLBM}, materials science \cite{PhysRevE.110.025301}, and biology \cite{PhysRevResearch.5.043096}, and has been widely adopted by research groups and corporations. Despite its success, the computational limitations of the method for solving industrially relevant systems, comprising a large number of computational points (lattice sites in the LBM context) and time steps, were found to be unavoidable. This problem is not particular to LBM, but rather to current classical methods for modelling partial differential equations. 

In light of the facts mentioned above, the emergence of quantum computing has given rise to new quantum algorithms that adapt LBM. The first quantum algorithm, applied quantum LBM (QLBM), was used to solve simple linear problems such as advection-diffusion \cite{Budinski_2021}. In this work, the authors reported, for the first time, an algorithmic advantage of quantum LBM over its classical counterparts. Their method included the collision, streaming, and macroscopic evaluation steps, scaling with the number of lattice sites as $O(\log^2 N)$. Still, the approach was restricted to a single time step without measurements, and both the state preparation and the measurement stage required circuits with linear depth O(N), which prevented these components from contributing to the claimed speedup. Shortly thereafter, a quantum algorithm for the Navier–Stokes equations based on the stream–vorticity representation was introduced \cite{Budinski2022QuantumNS}, providing circuits capable of reproducing LBM-type nonlinear dynamics. However, the original bottlenecks remained unaddressed. Over the past few years, interest in QLBM has expanded considerably, and the number of related publications has grown. Examples include new approaches for result extraction \cite{schalkers2024momentum}, initiatives toward open-source toolchains \cite{GEORGESCU2025109699}, 3D simulations of more complex flow configurations \cite{xiao2025quantum}, multi-step evolution for linear cases using dynamical circuits \cite{wawrzyniak2025dynamic} and a novel one-step simplified QLBM approach enabling a real-device demonstration of a Navier-Stokes flow with an immersed obstacle \cite{bastida2026quantumlbm}.

The combination of nonlinear terms and time-step evolution without intermediary measurements has been a challenge until now, and the main focus of current state-of-the-art research. In 2023, a first article attempted to address this challenge using Carleman linearization \cite{Sanavio2024LatticeBoltzmannCarleman}, a technique that has been successfully applied in other quantum algorithms to linearise equations by introducing additional unknown variables. Successive work focused on this issue, improving the efficiency of the algorithm but reducing the accuracy, which translated into a poor improvement overall \cite{Sanavio2024CarlemanGrad}. Nonetheless, these advancements were insufficient due to the high number of ancilla qubits and a very low success probability, which still did not enable the practical implementation of multiple time steps. To solve this, two main approaches have been proposed. The first approach utilises a novel time-preserving encoding that helps the local construction of the collision operator for Carleman linearization \cite{BastidaZamora2025QLBM}. This allows obtaining a lower depth and a higher success probability without sacrificing accuracy. However, the method relies on implementing non-unitary gates using a block-encoding technique known as the linear combination of unitaries (LCU) \cite{childs2012LCU}, which is impractical for many time steps. The second approach combines Carleman linearization with the warped phase transformation (WPT), introducing an additional variable that allows rewriting the nonlinear equation as a linear Schrödinger equation with a unitary operator. Nonetheless, its implementation requires a large number of qubits and circuit depth to implement Hamiltonian unitaries and the additional variables.  

Given the difficulty of the task, a new and promising idea soon emerged. Variational quantum circuits, where a set of gates depending on parameters is modified using a hybrid quantum-classical architecture following the principles of quantum machine learning (QML) \cite{cerezo2021VQA}, have been successfully applied to a wide range of technical challenges, including classification \cite{PhysRevResearch.7.013082}, quantum state preparation \cite{kerppo2025minimizingEntanglement}, optimisation \cite{PhysRevResearch.7.023141}, and others. Combining the capabilities of these circuits and previously known efforts to train the collision operator of LBM with classical neural networks (NN) \cite{corbetta2023LBcollision}, new studies related to variational quantum circuits (VQC) for QLBM collision operator were developed \cite{lactatus2025surrogate,Itani2025QMLLBM}. In these works, a VQC is trained to predict the collision operator of LBM. However, these do not directly address the problem of predicting nonlinear terms over multiple time-steps, where both phases and amplitudes are estimated. 

In this article, we give further insights into the problem of training QVC for the collision operator of QLBM. Specifically, we focus on generating nonlinear collision functions using the post-linear-collision distribution functions as input for the variational circuit, where only linear terms in $u$ have been employed. Additionally, we present novel results for simulations with more than one time step without intermediary measurements and high-accuracy solutions for more than one time step with and without intermediary measurements. The goal of the article is therefore to pave the way for understanding the strengths and limitations of using quantum machine learning (QML) for learning nonlinear physical processes, specifically the LBM collision operator.

The article is organised as follows. In Sec~\ref{sec:sec2} we present the relevant background concepts, such as LBM, its equivalences, and conservation laws, for a physically informed neural network and its corresponding quantum circuit. Sec~\ref{sec:sec3} introduces the two models proposed in this article: The R1 model (Subsec~\ref{sec:R1}), with one quantum register, aiming for simulations of several time-steps and nonlinearities and the R2 model (Subsec~\ref{sec:R2}), with two quantum registers, which shows high accuracy but is restricted with the need of measurements for each time-step. Finally, we present the conclusions of this work in Sec~\ref{sec:conclusions}.

\section{Known concepts: Lattice Boltzmann method and collision operator training}
\label{sec:sec2}
In this section, we will briefly introduce the LBM theory needed to follow this article and previous research on the training of the collision operator of LBM using quantum variational circuits.  This will serve as the foundation for the next section, where we will present novel analysis and results with a similar technique. Alternatively, readers can refer to the original articles addressing this problem \cite{lactatus2025surrogate,corbetta2023LBcollision} and to an extended introduction to LBM \cite{wagner2008practical}. 
\subsection{Lattice Boltzmann method}

\begin{figure}[htbp]
    \centering
    \includegraphics[width=0.35\textwidth]{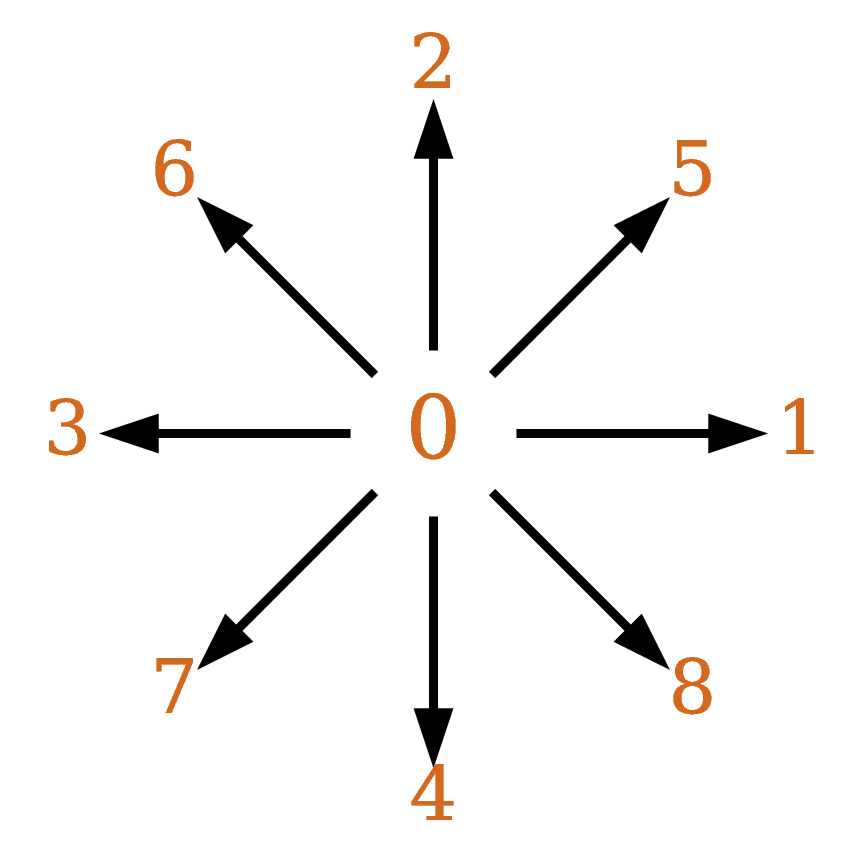}

    \caption{Scheme of channels for D2Q9. Each number represents the index used to map each channel.} 
    \label{fig:d2q9}
\end{figure}

LBM can be thought of as a family of models rather than a single computational method, as it has adopted many shapes and different implementations since its origin. One of the most widely used LBM is the Bhatnagar-Gross-Krook (BGK) approximation with a D2Q9 scheme, as shown in Fig~\ref{fig:d2q9}. 
The first number refers to the model's dimensions, while the second refers to the number of channels. The channels are the different directions in which a particle can be in each lattice site (computational point). In LBM, instead of computing the position of each particle, a probability distribution function $f_i(x)$ is used, with $i$ the propagation direction and $x$ the lattice site. Under this approximation, LBM can be mathematically expressed as
\begin{equation}
    f^{ref}(x+v,t+1)=f^{str}(x,t)+\Omega
    \label{eq:basic}
\end{equation}
with $\Omega$ the collision operator, which is in charge of modifying the probability distributions at each lattice site. After that, each $f_i(x)$ is shifted to $f_i(x+v_i)$ with $v_i$ the propagation direction of the channel. Using the BGK approximation, the elements of the $\Omega$ operator can be written as
\begin{equation}
\Omega_i=\frac{1}{\tau} \left(f_i^{eq}(x,t)-f_i^{str}(x,t)\right)
\label{eq:omega}
\end{equation}
where $f_i^{eq}$ is the equilibrium distribution function to which each $f_i$ converges.
To clarify the notation, from now on until the end of the paper, we will use $f$ without a superscript for any generic distribution function, and the following superscripts for the distribution functions at each of the following computational steps:
\begin{equation}
\begin{aligned}
    f^{\text{str}} &: \text{Post-propagation distribution,} \\
    f^{\text{lin}} &: \text{Post-linear collision distribution,} \\
    f^{\text{vqc}} &: \text{Predicted distribution by VQC (to be introduced),} \\
    f^{\text{ref}} &: \text{Analytical nonlinear post-collision distribution,} \\
    f^{\text{eq}}  &: \text{Equilibrium distribution } (f^{\text{eq}}=f^{\text{ref}} \text{ if } \tau=1).
\end{aligned}
\label{eq:notation_definitions}
\end{equation}
The logical progression of these states through our proposed collision operator can therefore be summarised as:
\begin{equation}
    f_i^{\text{str}} \xrightarrow{\text{Linear operator } A} f_i^{\text{lin}} \xrightarrow{\text{Quantum circuit } U(\theta)} f_i^{\text{vqc}} \approx f_i^{\text{ref}}
    \label{eq:notation_pipeline}
\end{equation}

Therefore, LBM captures weakly non-equilibrium dynamics. Although deriving the equilibrium distribution requires a detailed introduction, its final expression is relatively simple.
\begin{equation}
f_i^{eq}(x,t)=w_i \rho(x,t) \left(1+\frac{u c_i}{c_s^2}+\frac{(uc_i)^2}{2c_s^4}-\frac{u u}{2c_s^2}\right)
\label{eq:f_eq}
\end{equation}
with $w_i$ the index weight, which depends on the propagation channel, $\rho(x)$ the density at the lattice $x$, $u$ the local velocity and $c_i$ the direction of the propagation. For D2Q9, the velocity can be written as 
\begin{equation}
\begin{aligned}
u_x &= f_1 + f_5 + f_8 - f_3 - f_6 - f_7, \\
u_y &= f_2 + f_5 + f_6 - f_4 - f_7 - f_8.
\end{aligned}
\end{equation}
according to Fig \ref{fig:d2q9}. We can rewrite \eqref{eq:f_eq} as
\begin{equation}
   f_{i}^{eq}=w_i \rho \left(1+Af_i+Bf_i^2\right)  
   \label{eq:f_eq_orders}
\end{equation}
with $A f_i\gg B f_i^2$ as it is an approximation at second order. Using \eqref{eq:basic}, \eqref{eq:omega} with $\tau=1$ and $\rho=1$ to simplify, and \eqref{eq:f_eq_orders}, we obtain
\begin{equation}
    f_i(x+v_i,t+1)=w_i \left(1+Af_i(x)+Bf_i^2(x)\right)
\end{equation}
As we see, the final equation depends only on two non-constant variables, $f_i(x)$ and $f_i^2(x)$. This expression is the simplest form of LBM and is sufficient for a large number of simulation cases where density fluctuations are slight, and the method is stable.

\subsection{Concepts to learn the collision operator of LBM}
\label{sec:concepts_ML_collision}
Among the diverse studies previously published, training the collision operator of LBM using machine learning and QLBM using quantum machine learning, three articles stand out.   The first article in 2023 by Corbetta \emph{et al.} addressed this problem using a classical neural network \cite{corbetta2023LBcollision}. The second article directly addresses the training of the collision operator of LBM in a quantum circuit with a unitary operator via a variational quantum circuit \cite{lactatus2025surrogate}. Finally, a recent article by Itani \emph{et al.} develops a new statevector encoding using a similar architecture as previous articles. The authors conclude that the model performs poorly at high velocities and is not ideal for practical purposes. First, we will introduce the most important concepts from the first work. The first concept is collision invariants and equivariance, which are quantities that must be conserved due to the physical constraints of LBM. Implementing these concepts can enhance results and yield better solutions. Here we list the equivariances that an ML collision must satisfy.

\begin{enumerate}
    \item \textbf{Scale equivariance:} Scaling the distribution functions $f_i$ with a factor $\lambda$ before applying the collision operator $\Omega$, must provide the same results as scaling after the collision operator 
    \begin{equation}
        \Omega(\lambda f_i)=\lambda \Omega(f_i)
    \end{equation}
    \item \textbf{Rotation and reflection equivariance:} Two-dimensional collisions are equivariant with respect to rotation and reflection. This means that the post-collision distribution function $f^{ref}_i$ produced only depends on the relative position of the initial distribution functions $f$, with independence of the distribution function direction noted with subindex $i$. This symmetry can be easily understood if we think about the distribution functions as particles. Due to rotational invariance, the post-collision momenta of two particles with incident momenta $\mathbf{p}_i$ and $\mathbf{p}_j$ do not depend on the absolute orientation of the system in the laboratory frame. Instead, the collision dynamics are determined exclusively by the initial momenta magnitudes and the relative angle $\theta_{ij}$ between the incident trajectories. When using a D2Q9 model, this independence corresponds to preserving the 8th order dihedral symmetry group of the lattice D8 with $r$ (90° rotation) and $s$ (mirroring) as generators of the group. The group is
    \begin{equation}
        D_8=\{I,r,r^2,r^3,s,rs,r^2s,r^3s\}
    \end{equation}
    with the invariance expressed as
    \begin{equation}
\Omega(\sigma f_i) = \sigma \Omega(f_i) \qquad \forall \sigma \in D_8    \end{equation}
    \item \textbf{Mass, momentum invariance and positivity:} Collisions between particles in fluid dynamics conserve mass and momentum. These can be expressed as
\begin{equation}
\begin{aligned}
    \sum_{i=0}^8 \left(f_i^{\text{ref}} - f_i^{\text{str}}\right) &= 0, \\
    \sum_{i=0}^8 \left(f_i^{\text{ref}} - f_i^{\text{str}}\right) c_i &= 0 \\
\end{aligned}
\end{equation}
Additionally, the final distribution functions must be positive 
\begin{equation}
f_i^{\text{ref}} > 0 \qquad \forall \, i \in [0,8]
\end{equation}
\end{enumerate}

After understanding all the properties the collision operator must preserve, the next question is what we need and how we will train the operator. As our focus is on using a quantum circuit for the training, which will be addressed in the following subsection, we will only revise the general concepts. The first element we need to train the collision operator is a dataset with our initial distribution function $f$ as input and our desired target as output, which allows the learned operator to be as general as possible. As we will see, a truly general nonlinear collision operator for a quantum computer across all ranges of velocities, densities, and distribution functions is not realistic. Depending on the simulation, the model may not generalise well across different initial conditions, obstacles, boundary conditions, or test cases. In this regard, convergence guarantees are not found, and the analysis is purely heuristic. In the worst-case scenario, a different collision operator will be trained for each test case with a slight variation of possible initial conditions. Then, the decision to create a dataset depends on our specific needs. The more general our dataset is, the worse the model will perform at adapting to each case in the dataset, while the more specific it is, the better the results will be for that particular data range. While \cite{corbetta2023LBcollision,lactatus2025surrogate,Itani2025QMLLBM} use an artificial dataset to train the model, we also obtained accurate results using datasets generated from real simulations. In general, we will use datasets composed of real simulated flows unless otherwise stated. However, for completeness, we will include the procedure for creating an artificial dataset that resembles the real simulation. 

\begin{enumerate}
    \item \textbf{Velocity distributions:} As we saw, the operators $A$ and $B$ depend on local velocities. To create an artificial dataset for our simulation, the first step is to determine the velocity range and its distribution. Another important factor is to set the correlation between $u_x$ and $u_y$ for a two-dimensional system. For example, for our test case (Taylor-Green vortex), both velocities are related to an angle $\theta$ such that 
    \begin{equation}
        u=\left(u_{x_0}cos(\theta),u_{y_0}sin(\theta)\right)
    \end{equation}
    with 
\begin{equation}
\begin{aligned}
    u_{x_0},\, u_{y_0} &\sim \mathcal{U}\!\left(0,\, u_0^{\max}\right), \\
    \theta &\sim \mathcal{U}\!\left(-\pi,\, \pi\right).
\end{aligned}
\end{equation}
with $\mathcal{U}(x,y)$ the uniform distribution between values $x$ and $y$.

In this case, we do not address the density because we will assume it's very similar at each lattice site. For quantum circuits, density will not be an issue, but this should be included otherwise. In Fig~\ref{fig:velocity_correlations}, we observe an example comparison of the correlations and magnitudes between a real simulation and an artificial dataset generated as stated. 

\begin{figure}[htbp]

    \subfloat[\textbf{Velocity correlation obtained from LBM simulation}]{
        \includegraphics[width=0.95\linewidth]{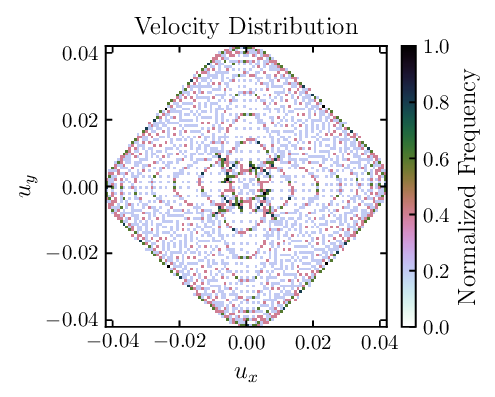}
    }
    \hfill
    \subfloat[\textbf{Artificial velocity correlation}]{
        \includegraphics[width=0.95\linewidth]{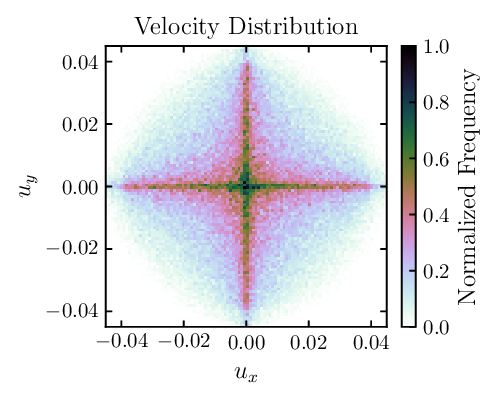}
    }

    \caption{Comparison between the magnitudes and correlations of $u_x$ and $u_y$ of a Taylor-Green vortex simulation and the generated dataset with described assumptions. The simulation parameters are $N_x=N_y=64$ lattice sites, $T=50$ time steps, velocity $u_0=0.05$, relaxation $\tau=1$} and $k_x=k_y=\pi/N_x$ wavenumber. 
    \label{fig:velocity_correlations}
\end{figure}

\item \textbf{Equilibrium distributions:} With the computed velocities, we calculate the equilibrium distributions following \eqref{eq:f_eq}

\item \textbf{Non-equilibrium distributions:} 
At each time step, $f_i$ will be weakly out of equilibrium at the beginning of the collision as the propagation step generates non-equilibrium distributions even for the case of $\tau=1$. Knowing the distribution and magnitude of the non-equilibrium dynamics produced by the propagation step will be important. In general, the non-equilibrium probability distributions can be computed using a normal distribution with variance $\sigma$ sampled from an interval. The maximum values of the standard deviation can be estimated, extrapolated, or exactly computed with one or several time-steps of our simulation case. The standard deviation will change over the course of the simulation. It will be maximal at the beginning of the simulation for flows with steady state solutions, or will remain roughly unchanged for systems without a steady state. In any case, this has to be estimated with a greater or lower precision depending on the generalisation capabilities we want for our model. We can express this as
\begin{equation}
\begin{aligned}
    \sigma_i &\sim \mathcal{U}\!\left(\sigma_i^{\min},\, \sigma_i^{\max}\right), \\
    f_i^{neq} &\sim \mathcal{N}\!\left(0,\, \sigma_i^{2}\right).
\end{aligned}
\end{equation}
with $\mathcal{N}(\mu,\sigma)$, the normal distribution with average $\mu$ and standard deviation $\sigma$.

The non-equilibrium distribution can be modified to be massless and momentumless. However, as the order of magnitude of the distribution is small, it will not modify the target distributions. The decision on how to create the dataset is purely technical, and it is to be designed and tested heuristically. In Fig~\ref{fig:feq_comparison}, we can observe an example comparison between the out-of-equilibrium contributions of an LBM simulation and an artificial dataset generated as stated. 

\begin{figure}[htbp]
    \centering
    
    \begin{subfigure}{0.48\textwidth}
        \centering
        \includegraphics[width=\linewidth]{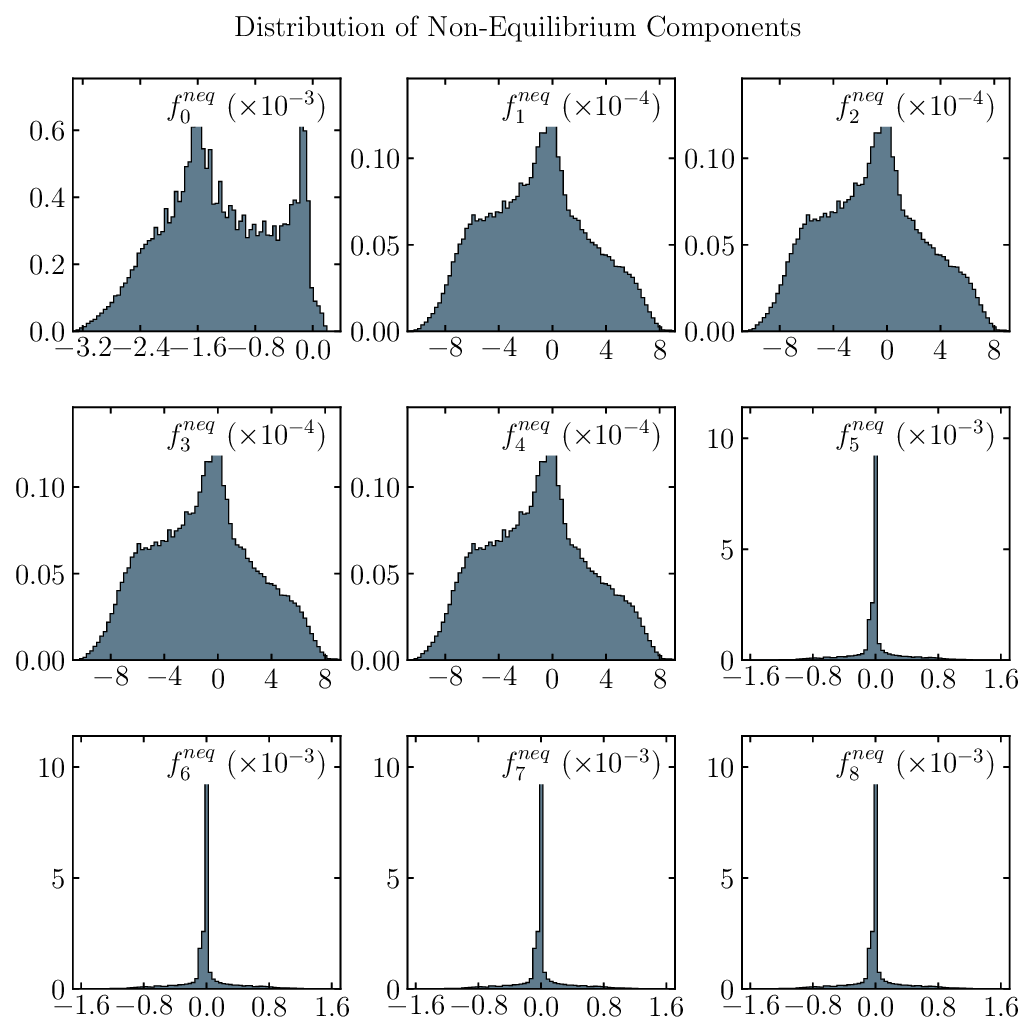}
        \caption{Using LBM simulation }
        \label{fig:f_neq_dataset}
    \end{subfigure}
    \begin{subfigure}{0.48\textwidth}
        \centering
        \includegraphics[width=\linewidth]{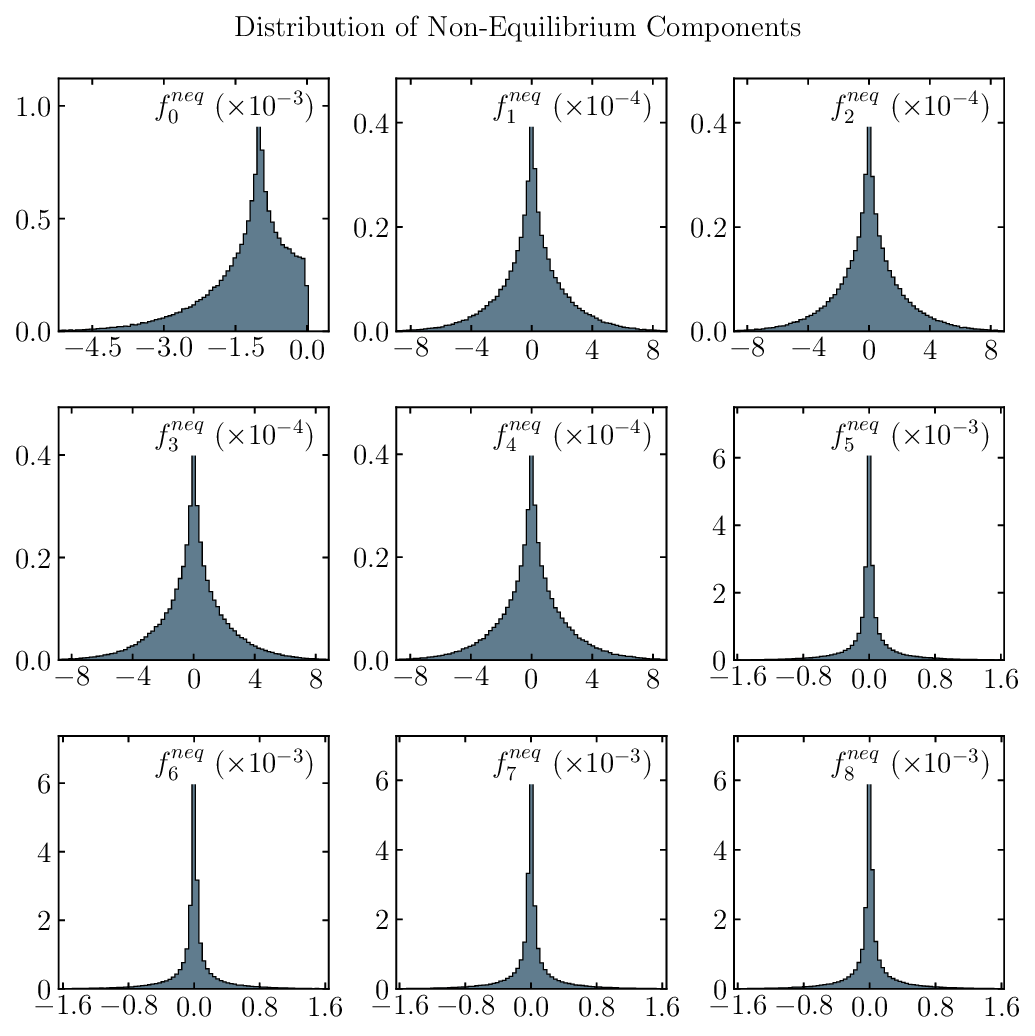}
        \caption{Using artificial dataset}
        \label{fig:f_neq}
    \end{subfigure}

    \caption{Comparison of the out-of-equilibrium contributions of each distribution function from the Taylor–Green vortex simulation with those obtained from the artificial dataset. The simulation parameters are $N_x=N_y=64$ lattice sites, $T=50$ time steps, velocity $u_0=0.05$, relaxation $\tau=1$ and $k_x=k_y=\pi/N_x$ wavenumber (which means that only one vortex is considered during the simulation).}
    \label{fig:feq_comparison}
\end{figure}

\item \textbf{Final pre- and post-collision distributions:}
After obtaining the equilibrium distribution and the non-equilibrium deviations, we can compute our final initial (pre-collision) and final (post-collision) distributions as
\begin{equation}
\begin{aligned}
    f_{i}^{\text{str}} &= f_i^{\text{eq}} + f_i^{\text{neq}}, \\
    f_{i}^{\text{ref}} &= \Omega(f_i^{\text{str}}).
\end{aligned}
\end{equation}
\end{enumerate}

Notice that the distribution of $f_i^{\text{neq}}$ is not general and will be case dependent as previously introduced. The optimal artificial construction of the dataset using this procedure is a question for future research.

\subsection{Adapting previous methods using variational quantum circuits}
\label{sec:dihedral_d2q9}
Now we will focus on the translation of previous concepts to the quantum computing already  addressed in previous research by Lăcătuş \textit{et al} in \cite{lactatus2025surrogate}. A major distinction between classical neural networks and variational quantum circuits lies in the circuit architecture and the limitations inherent to the quantum setting. In contrast to its classical twin, in the quantum circuit, we do not have nonlinear activation functions, so all computation must be linear. Additionally, the goal in the quantum circuit is not to find a function that interpolates and extrapolates all our data, but a unitary operator that, after being applied, transforms the quantum state, density operator, measured probabilities or expectation values (depending on the cost function) to the target and allows its repeated computation efficiently. This subtle but important concept makes variational quantum circuits completely different from their classical counterparts. 

As introduced in previous work, the greatest difficulty in training LBM distribution functions is to maintain unitarity while preserving the group symmetry described in Sec .~$\ref {sec:concepts_ML_collision}$. A way to do this is to set a particular ansatz that preserves the constraint using unitary quantum gates. It is important to remember that an ansatz in quantum circuits is not more than a set of parametrised gates that set the degrees of freedom of our circuit. If the circuit does not have any gates between two qubits, then the optimisation will be restricted to find a total unitary without a direct entanglement between the two. How we define the ansatz will determine what unitary we obtain and how well we can approach our desired solution. To do this any quantum gate $U_{SQC}(\theta)$ in the circuit  must follow 
\begin{equation}
    U_{\sigma}U_{SQC}(\theta)U_{\sigma}^\dagger=U_{SQC}(\theta) :  \forall \sigma\in D_8
\end{equation}
or in other words, the operators must commute. Considering that general parametrised single-qubit gates combined with entangling two-qubit gates form a universal set, the task is to find which one and two-qubit gates to use. First, if we consider one qubit rotational gates defined by the Pauli matrices $s_i$, its general expression is
\begin{equation}
U_n(\theta)=\text{exp}\left(-i\frac{\theta}{2}s_n\right) : n\in[x,y,z]
\end{equation}
which commutes with $D_8$ if the same unitary $U_n$ is applied to each qubit representing the distribution functions. For the case of D2Q9, we will need four qubits for nine distributions $f_i$, the unitary to be used is
\begin{equation}
U_n(\theta)=\left(exp\left(-i\frac{\theta}{2}\sigma_n\right)\right)^{\otimes 4} : n\in[x,y,z]
\end{equation}
For two-qubit gates, we have that
\begin{equation}
U_{\sigma}W_E(\theta)U_{\sigma}^\dagger=W_E(\theta):  \forall \sigma\in D_8
\label{eq:two_1}
\end{equation}
following the notation described in \cite{lactatus2025surrogate} where $W_E(\theta)$ is an unitary obtained by applying two-qubit gates and $W_{ij}(\theta)$ is a two-qubit gate entangling $i$ and $j$ qubits. Using that $W(\theta) = \prod\limits_{(i,j)\in E} W_{ij}(\theta)$ with $E$ the set of possible pairs,
\begin{equation}
U_{\sigma} W_E(\theta) U_{\sigma}^\dagger
=
\prod_{(i,j) \in E} W_{(\sigma(i), \sigma(j))}(\theta)
\quad \forall \, \sigma \in D_8
\label{eq:two_2}
\end{equation}
Combining \eqref{eq:two_1} amd \eqref{eq:two_2},
\begin{equation}
\prod_{(i,j)\in E} W_{ij}(\theta)=\prod_{(i,j) \in E} W_{(\sigma(i), \sigma(j))}(\theta) \quad \forall \, \sigma \in D_8
\end{equation}
Therefore, we will only use gates acting on the subset of elements $E_2 \subset E$, such that all pairs in $E_2$ form a set closed under the action of the $D_8$ group, with each gate parameterised by the same angle $\theta$. There are two index sets that fulfil both conditions
\begin{equation}
\begin{aligned}
E_{\text{axial}} &= \left\{ (0,1), (1,2), (2,3), (3,0) \right\}, \\
E_{\text{diag}}  &= \left\{ (0,2), (1,3) \right\}.
\end{aligned}
\end{equation}
Additionally, to ensure that the same angle is applied to each of the indices, we will use Ising layers along the $x$,$y$ or $z$ axis, which can be expressed, for example, for the case of $XX$ as
\begin{equation}
XX(\theta)=\prod_{(i,j)\in E} U_{ij}(\theta_{ij})
\end{equation}

\section{Variational quantum circuits for QLBM: Towards nonlinearities for multiple time-steps}
\label{sec:sec3}
The novelty of this article lies in the trainability of the LBM collision operator using variational quantum circuits. Unlike previous works, we focus on training the nonlinear terms, concatenating multiple time steps, and analysing the conditions under which both can be achieved. Two models are proposed: the one-register model, R1, and the two-register model, R2. All concepts previously known from previous research will be used as introduced in Sec.~$\ref{sec:sec2}$. For both cases, we will use a combination of Pennylane \cite{bergholm2018pennylane} and Jax for training the quantum circuits with Adam optimiser \cite{kingma2014adam} and the parameter-shift rule \cite{PhysRevA.98.032309} to calculate the gradients. For all benchmarks in this article, we will use $B=20$ layers (no observed improvements beyond this), divide each epoch into batches of 32 vector distribution functions, and use a learning rate $l_r=10^{-4}$ with the Adam optimiser. The training will use a hybrid architecture. The circuit will be encoded, executed, and measured using an emulated quantum circuit, and the gradients will be calculated and updated classically at each training iteration. In this article, we will focus on nonlinearities and time-step concatenation. This means that for all cases, we will assume the linear operator $A$ to be given, as the input distribution function will be the post-linear-collision $f^{lin}$. The operator $A$ can be either manually built, constructed using automatic gate decomposition procedures \cite{mottonen2006decompositions}, or learned via VQC.

\subsection{VQC for QLBM with one register: R1 model}
\label{sec:R1}
\begin{figure*}[t]
    \centering
    \includegraphics[width=1\textwidth]{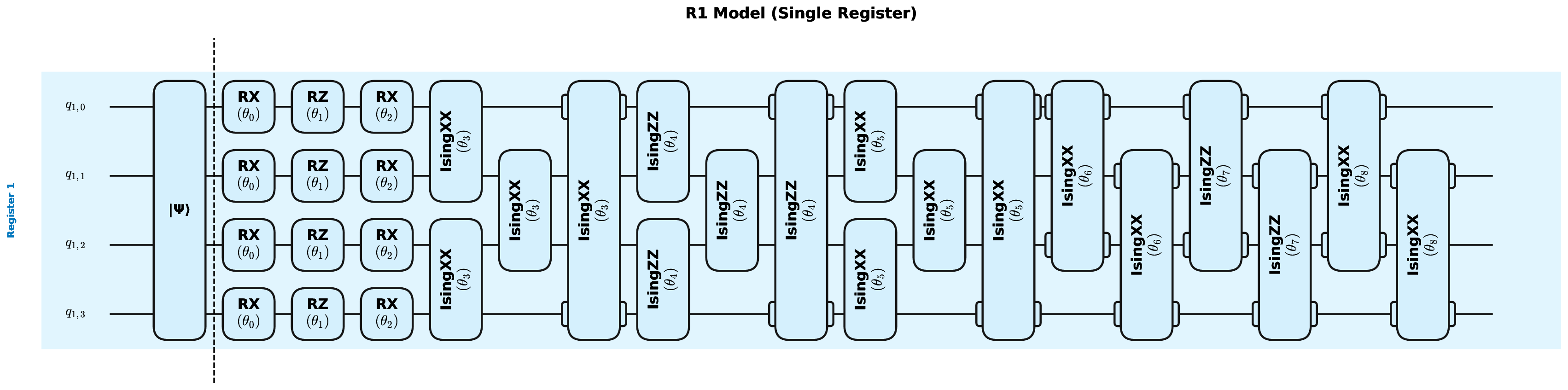}
    \caption{Schematic representation of the one-register Variational Quantum Circuit (R1 model) for the Quantum Lattice Boltzmann Method with one layer $B=1$. The architecture follows the same principles of previous research \cite{lactatus2025surrogate} to conserve dihedral symmetry as introduced in Sec~\ref{sec:dihedral_d2q9}. It also includes three different gates for each qubit, following the Euler angles principle for general rotations.}
    \label{fig:1r_circuit}
\end{figure*}

The simplest case for training a VQC for the LBM collision operator is the R1 model with one quantum register of four qubits, encoding nine distribution functions. This model ansatz follows the developments of Lăcătuş \textit{et al}. The quantum circuit used during training for one layer ($B=1$) can be seen in Fig~\ref{fig:1r_circuit}. We will divide this section into four subsections. In Subsec~\ref{subsec:R1_1}, we will introduce the model's encoding, ansatz, loss functions and early results. In Subsec~\ref{subsec:R1_2}, we will test how the model behaves and performs under different macroscopic conditions such as different maximum velocities $u_{max}$ and different relaxation parameter $\tau$. In Subsec~\ref{subsec:R1_3}, we will briefly discuss what happens when we modify the model to make it nonunitary and the advantages and disadvantages observed. Finally, once the model's capabilities and best practices are presented, we will present the results for different simulation cases in Subsec~\ref{subsec:R1_4}: the relaxation of an initial Kolmogorov flow, the vorticity field of a flow with a flat plate as an obstacle, and a forced flow creating orthogonal Gaussian jets. While we will not use it in the current tests, for the sake of simplicity of the procedure, we note that the methodology of  \cite{Skolik2021}, which shows that training the model layer by layer (adding layers during the simulation) improves results and mitigates barren plateaus, is a highly recommended approach to complement ours. Nonetheless, the effect is limited and better suited to fine-tuning. 

\subsubsection{Introducing the R1 model}
\label{subsec:R1_1}
As initially stated, we will first introduce the quantum state encoding, the ansatz, the loss function and the main methodology used. 
\begin{enumerate}
    \item \textbf{Quantum state encoding:} For the training the initial distribution functions used will be $f^{lin}=Af^{str}$. The initial statevector is
    \begin{equation}
        \ket{q_{1,i}}=\sum_{i=0}^8 \frac{\sqrt{f_i}}{\sqrt{\rho}} \ket{c_i}
    \end{equation}
    with $c_i$ using a special mapping to preserve $D_8$ group invariance. Each qubit will be associated to a primary direction: $f_0\rightarrow \ket{c_0}=\ket{0000}$, $f_1\rightarrow \ket{c_1}=\ket{0001}$,$f_2\rightarrow \ket{c_2}=\ket{0010}$,$f_3\rightarrow \ket{c_3}=\ket{0100}$,$f_3\rightarrow \ket{c_4}=\ket{1000}$  and the rest of distributions will be the sum of its constituents. As an example $f_5\rightarrow\ket{c_1+c_2}=\ket{0011}$. 
    \item \textbf{Quantum circuit ansatz:}
    The ansatz has to be general enough to be flexible, yet restrictive enough to avoid barren plateaus. Following previous work as introduced in Sec~\ref{sec:concepts_ML_collision}, we will use single qubit gates based on Pauli operators $R_x$ and $R_z$ and two-qubit gates, Ising gates $XX$ and $ZZ$, for entanglement. For the most general possible rotation in each axis, an Euler angles based approach is used, where each set of gates is divided into $X$, $Z$ and $X$ rotation in that order. Figure~\ref{fig:1r_circuit} shows the ansatz used in detail.
\item \textbf{Loss function:} The loss function will differ depending on our goal. According to our focus, we have two options: Using the mean square error (MSE) with the density operator as metric or a combination of the amplitude MSE and the phase, as seen in \eqref{eq:loss_amp_phase}, with $\lambda$ a parameter which changes the intensity of the phase compared to the amplitude. The error rate naturally accounts for the global phase by using a $mod$ function. 
\begin{equation}
\begin{aligned}
L_{A\phi}^{\lambda} &= (1-\lambda) \sum_i \left( |\Psi_i^{\rm vqc}|^2 - |\Psi_i^{\rm ref}|^2 \right)^2 \\
&\quad + \lambda \sum_i |\Psi_i^{\rm vqc}|^2 \frac{\left( \left( \Delta\Phi_i + \pi \right) \bmod 2\pi - \pi \right)^2}{4\pi^2}
\label{eq:loss_amp_phase}
\end{aligned}
\end{equation}
where $\Psi_i$ represents the complex amplitude of the $i$-th computational basis state, and the phase terms are extracted as $\Phi_i = \arg(\Psi_i)$. The term $\Delta\Phi_i$ accounts for the relative phase difference after aligning the global phase to the first basis state ($i=0$), defined as:
\begin{equation}
\Delta\Phi_i = \Phi_i^{\rm vqc} - \Phi_i^{\rm ref} - \left(\Phi_0^{\rm vqc} - \Phi_0^{\rm ref}\right)
\end{equation}
This formulation introduces two key improvements: it uses a convex combination controlled by $\lambda$ to balance the penalties, and it weights the phase loss by the predicted probability $|\Psi_i^{\rm vqc}|^2$, ensuring the optimiser does not waste resources correcting the phase of states with near-zero amplitude.

In general, we will use mean squared error (MSE) to compare the measured amplitudes and phases of each distribution function $f_i^{vqc}$ with the target $f_i^{ref}$ using our desired metric. Another alternative to target the amplitude and phases is to use the density operator, which rescales the phases to the amplitudes in the state vector. This means that the phase accuracy will be linked to the amplitudes. The higher the amplitude, the more important the phase accuracy becomes. We define this loss as
\begin{equation}
\begin{aligned}
L_{\rho} = \| \rho^{\mathrm{ref}} - \rho^{\mathrm{vqc}} \|_F^2
\label{eq:loss_rho}
\end{aligned}
\end{equation}
Additionally, to ensure that the model preserves the stress momenta $p_{xx}-p_{yy}$ and $p_{xy}$, the energy $E$, and the velocities $ u_x$ and $ u_y$, we can add terms targeting these macroscopic quantities. This way, we can define the macroscopic vector $M$ in terms of the distribution functions as 
\begin{equation}
M =
\begin{bmatrix}
u_x : f_1+f_5+f_8-f_3-f_6-f_7 \\
u_y : f_2+f_5+f_6-f_4-f_7-f_8\\
p_{xx}-p_{yy}: f_1+f_3-f_2-f_4 \\
p_{xy}: f_5+f_7-f_6-f_8 \\
E: -4f_0-f_1-f_2-f_3-f_4+2(f_5+f_6+f_7+f_8) \\
\end{bmatrix}
\end{equation}
\end{enumerate}
First, we will follow a naive approach. The main goal of this first analysis is to determine whether our loss function, dataset, and parameters are well-tuned, and whether we can obtain better results than LBM for $\tau=1$ than the linear model with only $O(u)$ terms in the equilibrium equation.
Regarding the loss function, we want to compare which is a better approach for time-step concatenation: The loss function $L_{A\phi}$ or $L_\rho$ targeting the amplitudes and phases of our measured quantum state $\ket{\Psi}^{vqc}$. For this, we will train the models using the distribution functions extracted from LBM simulation of a Kolmogorov flow with $\tau=1$, $u_{max}=0.05$ and test the maximum relative error in the velocity field after $T=50$ time-steps in a 2D Taylor-Green vortex (TGV) simulation with $L_x=64$ and same macroscopic conditions as the dataset as introduced in \cite{5bby-34zx}. In Table~\ref{tab:loss}, we report the maximum relative error in the velocity field after a given number of epochs for a dataset with $2\cdot 10^5$ elements. As we can see, the accuracy we can obtain with both methods is similar, but the usage of $L_\rho$ provides faster convergence. Notice that we used $\lambda=10^{-4}$ to set the relative weight of the phase and amplitudes, as it is the most accurate according to our tests. Despite this, we observed that separating the phase and amplitude in the loss function with a control parameter is superior for fine-tuning and achieving higher accuracy, even with longer training. 

\begin{table}[H]
\centering
\begin{tabular}{|c|c|c|}
\hline
 Epochs & $u$ relative error with $L_{A\phi}^{10^{-4}}$ & $u$ relative error with $L_{\rho}$ \\
\hline
5& 1 & 0.275\\
20 &0.4& 0.2 \\
40 & 0.32 & 0.14\\
60 & 0.14 & 0.14 \\
100 & 0.14 & 0.14\\
\hline
\end{tabular}
\caption{Comparison between the usage of $L_{A\phi}$ and $L_{\rho}$}
\label{tab:loss}
\end{table}

The next step is to think about the choice of the training dataset. In this regard, different approaches can provide good results. The first alternative is to build an artificial dataset in which each $f_i$ is distributed similarly to the target simulation, as shown in Fig~\ref{fig:f_neq_dataset}. Another alternative is to use data from a mix of simulations spanning the target velocity range and viscosity to ensure the training model is as generic and robust as possible. Finally, another alternative is to use data from the target simulation at a lower resolution to train our model and scale the number of sites on the quantum computer. As you may notice, for any of these approaches, we need to be able to simulate the target flow classically in the first place. However, given the large amount of data available from classical CFD and assuming the model is sufficiently robust, these concerns are unlikely to pose a problem. Given the difficulty of conducting a thorough analysis of the model's robustness and the limited bibliography on QML applied to LBM, our goal will be to provide insights into the model's capabilities. The analysis is conducted in a general fashion to understand and compare it with best practices, and we will not focus on the training dataset. Despite that, we think that it will be an interesting direction for future studies. If we compare all these approaches using $L_\rho$, we obtain the results displayed in Table~\ref{tab:datasets}. The results indicate that a dataset with a wide range of velocities and stress momenta, despite not being tuned for the target simulation, may be the superior choice here. In contrast, a fine-tuned dataset like the second or third can achieve high accuracy quickly at the cost of lower robustness. The results show that any of the approaches mentioned is suitable depending on our interests and also proves the model's robustness across different flows. 

\begin{table}[H]
\centering
\begin{tabular}{|c|c|c|c|}
\hline
 Epochs & Kolmogorov flow & Artificial & TGV - low resolution\\
\hline
5   & 0.275    & 0.14 & 0.08 \\
20  & 0.2  & 0.14   & 0.12 \\
40  & 0.14 & 0.06  & 0.16 \\
60  & 0.14 & 0.12  & 0.16 \\
100 & 0.14 & 0.175  & 0.10 \\
\hline
\end{tabular}
\caption{Comparison between different datasets for training}
\label{tab:datasets}
\end{table}

\begin{figure}[htbp]
    \centering
        \includegraphics[width=0.48\textwidth]{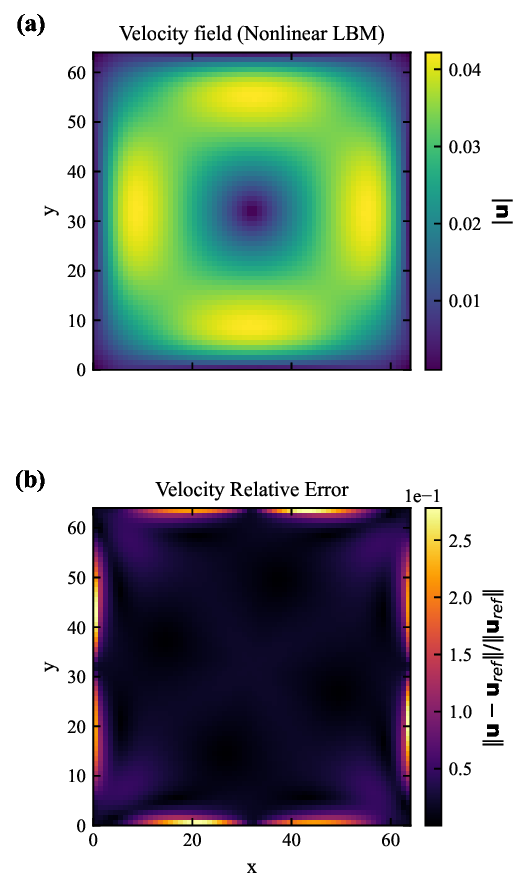}
    \caption{a) Velocity field for an LBM at second order $O(u^2)$ 2D TGV simulation with $u_{max}=0.05$ and $\tau=1$ with $L=64$ lattice sites per dimension and $T=50$ time-steps. b) Relative error between the velocity field of the nonlinear LBM and the linear LBM using only $O(u)$ to calculate the equilibrium distribution function. } 
    \label{fig:linear_comparison}
\end{figure}

Given the previous results and the comparison with Figure~\ref{fig:linear_comparison}, we can also draw a clear conclusion. It is possible to obtain a linear and unitary operator that, at least in certain flows, shows an advantage compared to the linear model. As mentioned before, this reduced relative error compared with the nonlinear case applies to a simulation in which we have carried the full statevector in an emulator rather than only the amplitudes, with the only assumption being an incompressible flow (same amplitude for every lattice site). While this is not a limitation of the model, it further simplifies the analysis.

\begin{figure}[htbp]
    \centering
        \includegraphics[width=0.48\textwidth]{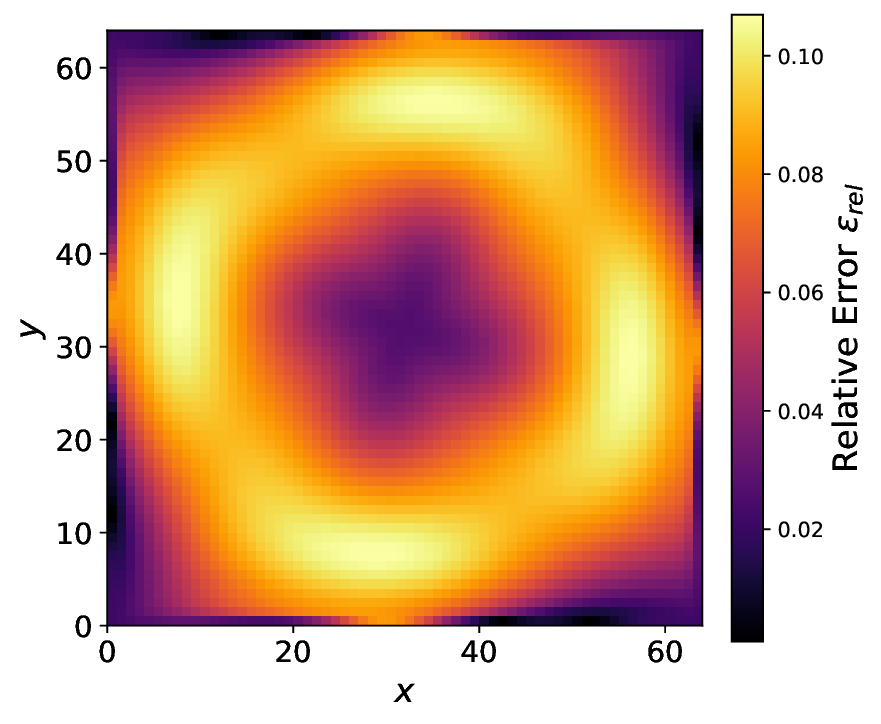}
    \caption{Relative error $\epsilon_{rel}$ of a TGV simulation using a QML model. The simulation parameters are: $64\times64$ lattice sites, $\tau=1$, $T=50$ time steps and $k_x=\frac{\pi}{N_X}$ and $u_{max}=0.05$. The total collision operator is composed of a first step with the linear collision operator and a second step involving the learned unitary from the QML model using $L_{\rho}$ as cost function.} 
    \label{fig:1r_re_rho}
\end{figure}

Using $L_\rho$ with a TGV at low resolution (16x16) lattice sites and trained during 100 epochs, we obtain the relative error in the velocity seen in Figure~\ref{fig:1r_re_rho}. The result is a higher accuracy in the boundary region (where higher relative errors were observed), but higher error in the rest of the domain. The maximum relative is 2.5 times lower than for the linear case, and the average relative error of the velocity field remains slightly lower. Overall, we obtain a better result than the linear case. 

\subsubsection{Dependencies of the model}
\label{subsec:R1_2}

Another essential aspect to analyse when considering a quantum machine learning model for LBM is to compare the performance of the trained unitary for different flows. We will focus on comparing the model at different velocities and viscosities. For this purpose, we will use Taylor-Green vortex simulation data at a lower resolution to directly train the model.

\begin{figure}[htbp]
    \centering
        \includegraphics[width=0.48\textwidth]{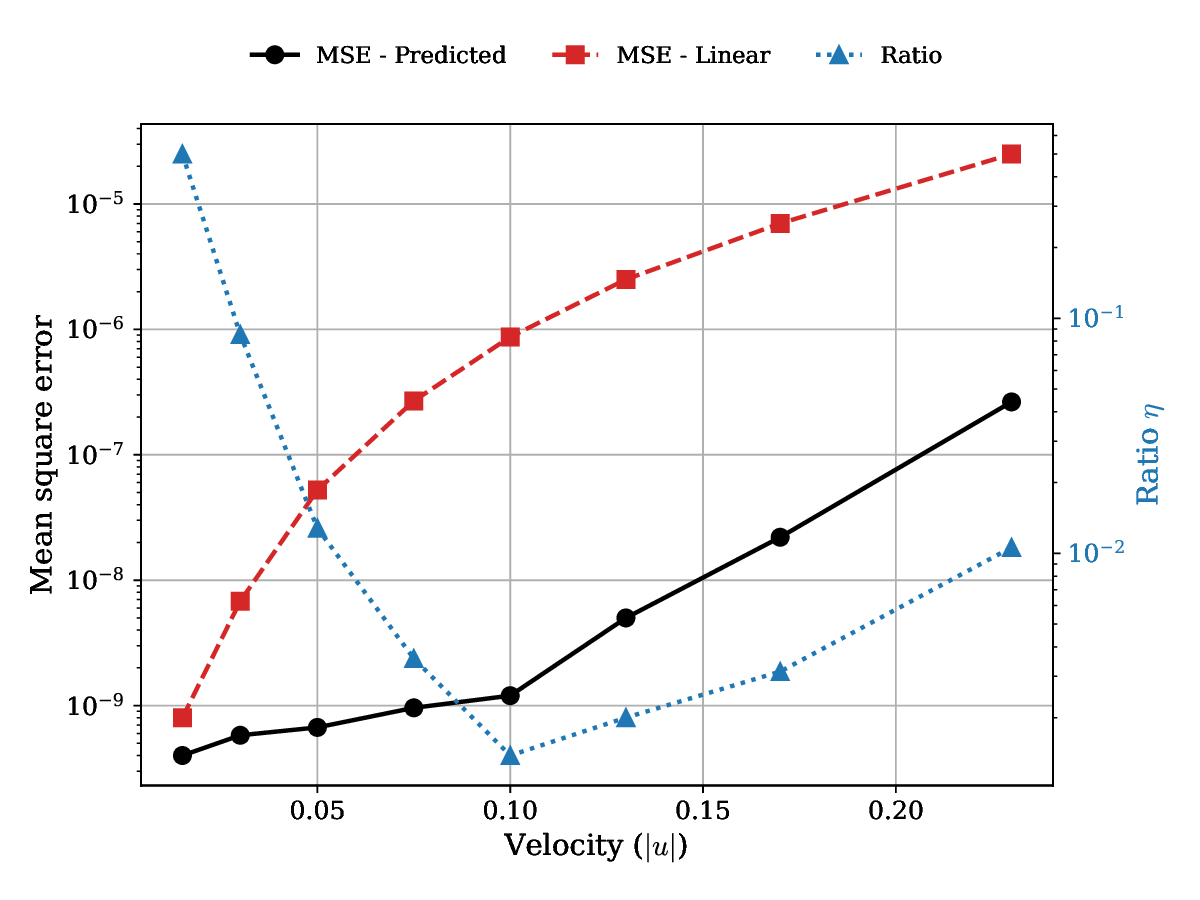}
    \caption{Comparison of the MSE of the predicted distribution functions $|f_i^{vqc}|$ with a QML when targeting the post-nonlinear-collision distribution $|f_i^{ref}|$ , compared with the mean square error between $|f_i^{ref}|$ and $|f_i^{lin}|$. The training used the data from 50 time-steps with different maximum velocities from a Taylor-Green vortex simulation. The ratio between the predicted MSE and the MSE with respect to the linear distribution is also shown. The cost function used by the QML is $L=10^{-4}L_{\phi}+L_{A}$, with each loss representing the phase and amplitude MSE, respectively, from \eqref{eq:loss_amp_phase}.}
    \label{fig:nonlinear_different_u}
\end{figure}

First, we will focus on TGV flows with different velocities and only on the average MSE of the distribution functions $f_i^{vqc}$. Here, velocity refers to the flow's maximum velocity. In Fig~\ref{fig:nonlinear_different_u}, we have displayed the results when we use $f_i^{lin}$ (post-linear-collision distribution functions using only $O(u)$ terms) as input and target $f_i^{ref}$ (post-collision LBM at second order). In the graph, two metrics are used for comparison. First, the MSE of the predicted distribution function $f_i^{vqc}$ (distribution function predicted by the model) compared with $f_i^{ref}$ and the MSE between $f_i^{lin}$ and $f_i^{ref}$. Second, we use the proportion between the nonlinear error and the linear error $\eta=\sum\limits_i\frac{(f_i^{vqc}-f_i^{ref})^2}{ (f^{lin}_i-f^{ref}_i)^2}$. In this case, the relative error of the distribution functions only measures the amplitude (the classical MSE). However, we use $L_{A\phi}$ as the loss function during training. In the figure, we can observe how, despite increasing the maximum velocity $u$, the ratio $\eta$ decreases up to $u=0.1$. For values beyond $u=0.1$, we obtain very high error rates and the ratio $\eta$ increases. This suggests that the VQC applies only to flows with low Mach numbers. Another reason for this may be the instability of the underlying BGK LBM model at higher speeds. Later, we will see how this trend is inverse when considering the MSE not only from the amplitudes of the statevector but also the phases, velocities and stress moments. 

\begin{figure}[htbp]
    \centering
        \includegraphics[width=0.48\textwidth]{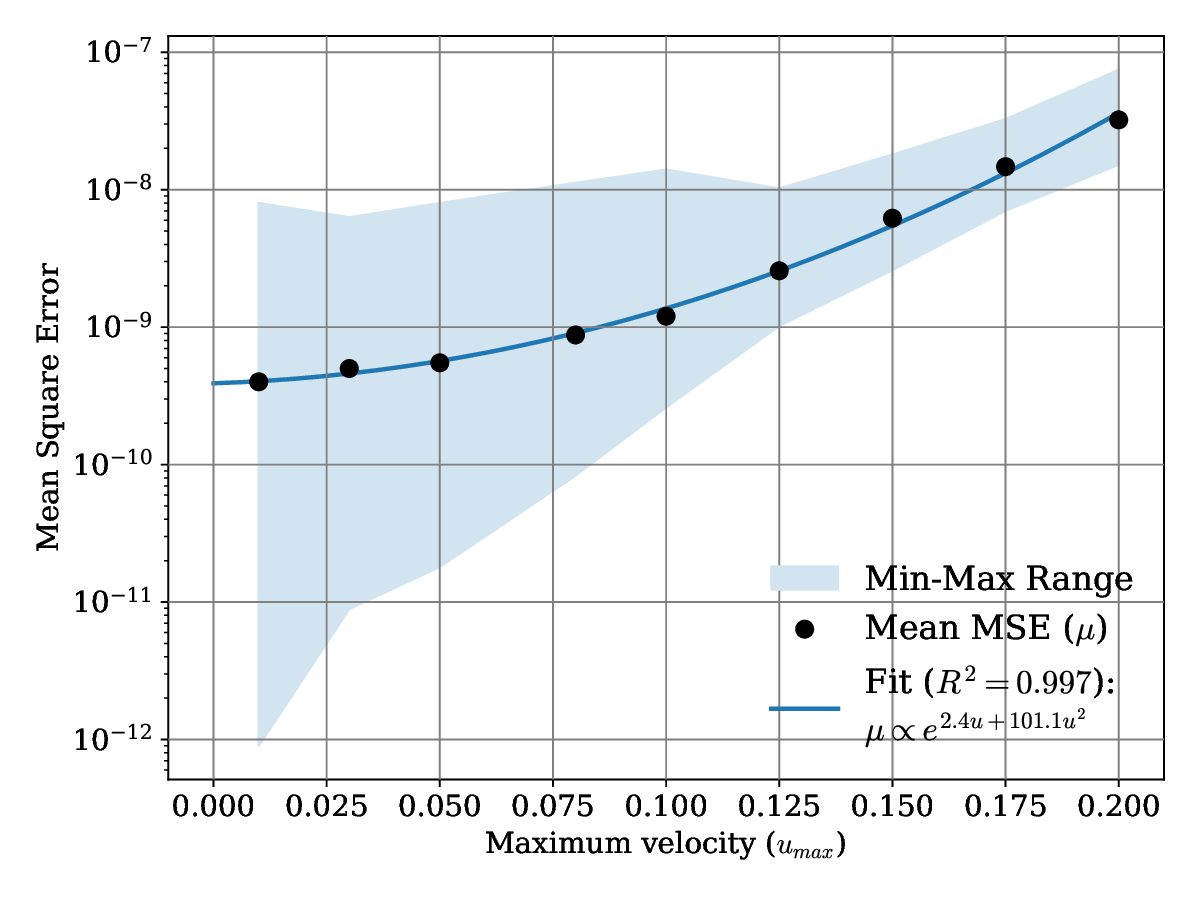}
    \caption{Mean square error of the predicted distribution functions $|f_i^{vqc}|$ using a quantum machine learning approach compared to $|f_i^{ref}|$. The training used the data from 50 time-steps with different maximum velocities of a Taylor-Green vortex simulation. The cost function of this experiment was $L=10^{-4}L_{\phi}+L_{A}$, with each loss representing the phase and amplitude MSE, respectively, from \eqref{eq:loss_amp_phase}. The dark blue line represents a polynomial fit as $c\cdot e^{a\cdot u+ b \cdot u^2}$ with $R^2$ the coefficient of determination.}
    \label{fig:nonlinear_fit_u}
\end{figure}

If we train the model for a larger number of epochs to improve accuracy (Fig~\ref{fig:nonlinear_fit_u}) and examine the maximum, median, and minimum errors for each maximum velocity $u_{max}$ during training, we can observe two phenomena. On the one hand, the average error increases with velocity as $O(e^{u^2})$ with very high fidelity $R^2=0.997$ up to $u=0.2$. On the other hand, the MSE range is reduced on a logarithmic scale. This may be due to a higher velocity range in the training dataset, along with a lower number of data points at lower velocities. In this case, we again observe that velocities above $u_{max}=0.1$ are not suitable for the model, as the maximum MSE quickly increases for $u_{max}>0.1$.

\begin{figure}[htbp]
    \centering
        \includegraphics[width=0.48\textwidth]{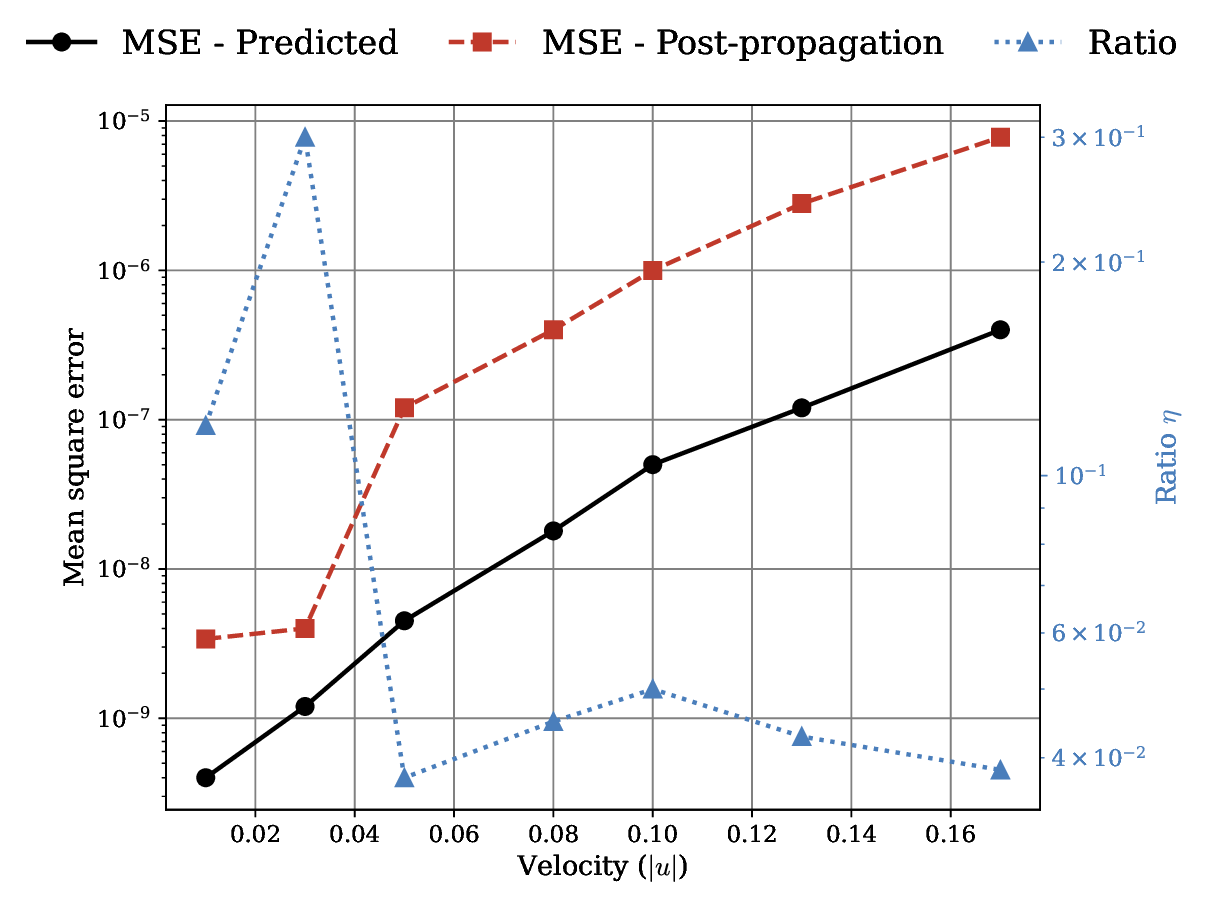}
    \caption{Comparison of the MSE of the predicted distribution functions $|f_i^{vqc}|$ with a QML when targeting the post-linear-collision distribution $|f_i^{lin}|$ , compared with the mean square error between $|f_i^{lin}|$ and $|f_i^{str}|$. The training used the data from 50 time-steps with different maximum velocities of a Taylor-Green vortex simulation. The ratio between the predicted MSE and the MSE with respect to the linear distribution is also shown. The cost function used by the QML is $L=10^{-4}L_{\phi}+L_{A}$, with each loss representing the phase and amplitude MSE, respectively, from \eqref{eq:loss_amp_phase}.}
    \label{fig:linear_different_u}
\end{figure}

The case of predicting the linear collision operator from the post-propagation distribution functions $f^{str}_i$ is different. As we can see in the Fig~\ref{fig:linear_different_u}, the ratio $\eta$ is constant in $u$ except for lower velocities and much higher than in the previous case. The post-collision distribution MSE increases exponentially as $(f_i^{str}-f_i^{lin})^2\propto e^u$. 

\begin{figure}[htbp]
    \centering
        \includegraphics[width=0.48\textwidth]{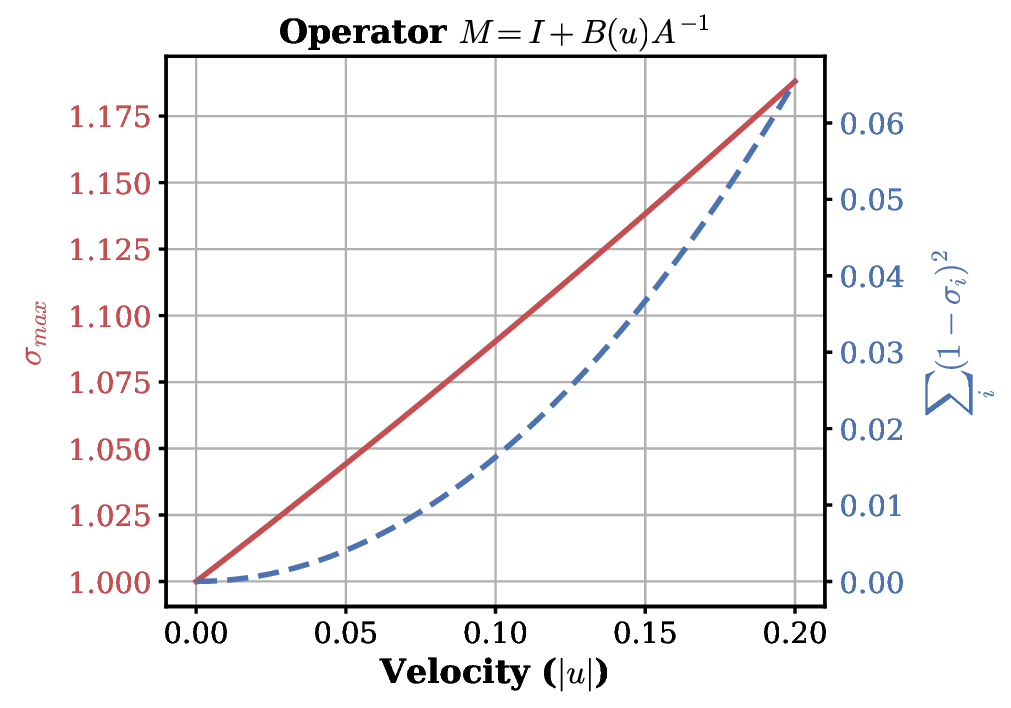}
    \caption{Nonunitary measure and singular value of $I+B(u)A^{-1}$, which describes the linear transformation from $f_i^{lin}$ to $f_i^{ref}$}
    \label{fig:spread}
\end{figure}

The quadratic growth of the predicted MSE in logarithmic scale in Figure~\ref{fig:nonlinear_different_u} for the nonlinear collision and the one targeting the linear collision in Figure~\ref{fig:linear_different_u}, is due to a combination of non-linearity and non-unitarity. This can be seen by analysing the change in the gap between the maximum and minimum eigenvalues for each case. If we decompose the collision operators as
\begin{equation}
\begin{split}
f_i^{\rm lin}    &= A f_i^{str} \\
f_i^{\rm ref} &= A f_i^{str} + B(u) f_i^{str}
\end{split}
\end{equation}
we obtain
\begin{equation}
f_i^{\rm ref}    = (I+B(u)A^{-1}) f_i^{lin} 
\end{equation}
Given the unitary operator $I + B(u) A^{-1}$, the degree of non-unitarity can be quantified as 
\[
\sum_i (1 - \sigma_i)^2,
\] 
where $\sigma_i$ are the singular values of the operator. As shown in Fig.~\ref{fig:spread}, the maximum singular value exhibits a linear dependence on the velocity. This indicates that higher velocities lead to a greater deviation from unitarity. The non-unitarity metric displayed in blue in the right axis corresponds to the MSE between the ideal unitary singular values (which are all 1) and the actual singular values of the operator $\sigma_i$. This justifies the quadratic mean square error with the velocity observed in Fig~\ref{fig:nonlinear_fit_u}, as the absolute difference between the input and output distributions increases linearly as the nonunitary does. The low non-unitarity supports the reason we obtain much higher accuracies when targeting $f_i^{ref}$ from $f_i^{lin}$ while keeping the operator unitary, rather than with $f_i^{lin}$ from $f_i^{str}$, as in the linear case, this metric is 6.5. Especially at low velocities, the model can linearly approximate the distribution functions because the target operator is nearly unitary. We believe this finding supports the idea of using a nonunitary operator with amplitudes diffusing into the undesired basis encoding in the linear case, and the use of the VQC unitary architecture specifically for the nonlinear case. This also reduces the problem's complexity and allows the model to approximate the nonlinear terms. In the next subsection, we will see that, despite that, relaxing the unitarity condition of the operator for the VQC may be helpful. 

Until now, we have compared the model's ability to predict the post-collision distribution functions $f^{ref}$ close to the analytical post-collision nonlinear distributions using MSE. However, in most cases, the main source of errors originating from the learned unitary will not be in the distributions but in the lack of momentum conservation. The lack of momentum conservation stems from insufficient decimal precision and from errors introduced by linear approximations of the nonlinear terms over a wide range of velocities. As we know, exact momentum conservation cannot be enforced during training without affecting the model's learning capability. In fact, one could train the model in momentum space, isolating the velocity and changing only the non-conserved magnitudes. However, in such architecture, after several attempts, we discovered that the model lacks predictive capabilities, as the velocity information is needed during the collision. Applying the unitary operator to a state containing velocity information will always yield imperfect conservation, as the entanglement between different momenta in the momentum space will modify the phase or amplitude of the velocity. To better understand these limitations, we will compare how velocity conservation changes as the maximum velocity increases and the inverse relationship with the prediction of the correct $f_i^{ref}$. 

\begin{figure}[htbp]
    \centering
        \includegraphics[width=0.48\textwidth]{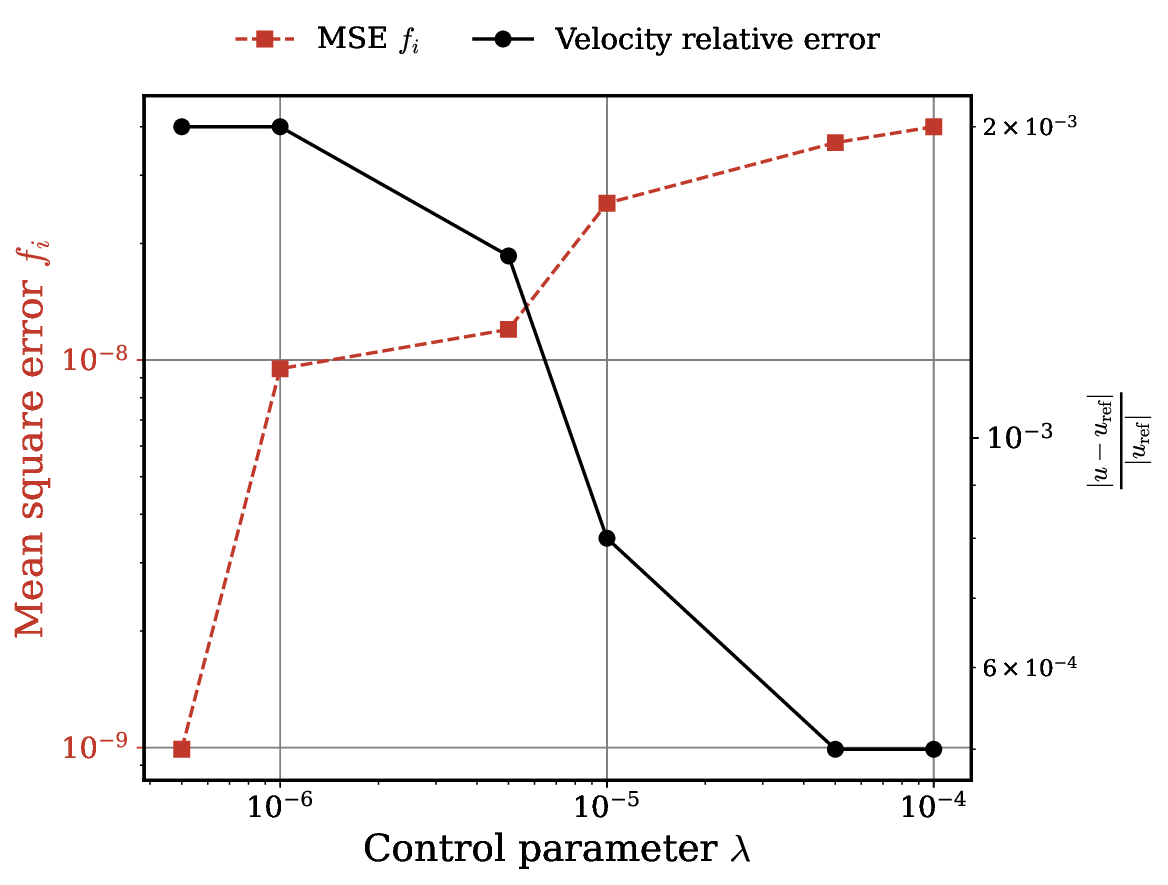}
    \caption{Comparison between the MSE between $f_i^{vqc}$ and $f_i^{ref}$ and the relative error of the velocities when using $L=L_{A\phi}+\lambda_u L_u$ as cost function, with $L_u$ representing the MSE between the predicted and analytical relative error of the velocity.}
    \label{fig:fi_vs_u}
\end{figure}

As seen in Fig~\ref{fig:fi_vs_u}, when we increase the weight of the control parameter $\lambda_u$ to increase the relevance of the velocity relative error, the $f^{vqc}$ prediction worsens. What we see is that the model can learn the distribution function $f^{vqc}$ and the final velocity field $u$ to some extent. However, if we force low relative error in the velocity field during training, the solution will converge to the initial distribution $f^{lin}$, favouring momentum conservation. We can conclude from this comparison that the lack of momentum conservation is a strong requirement for approximating the distribution functions with nonlinear contributions while maintaining the operator's unitarity. In practice, we recommend simply using the velocity MSE with $\lambda_u$ instead of using the relative velocity error. However, no large difference in results is observed when adding the velocity component, unless fine-tuning is performed for specific flows, as observed in our tests.

\begin{figure}[htbp]
    \centering
        \includegraphics[width=0.48\textwidth]{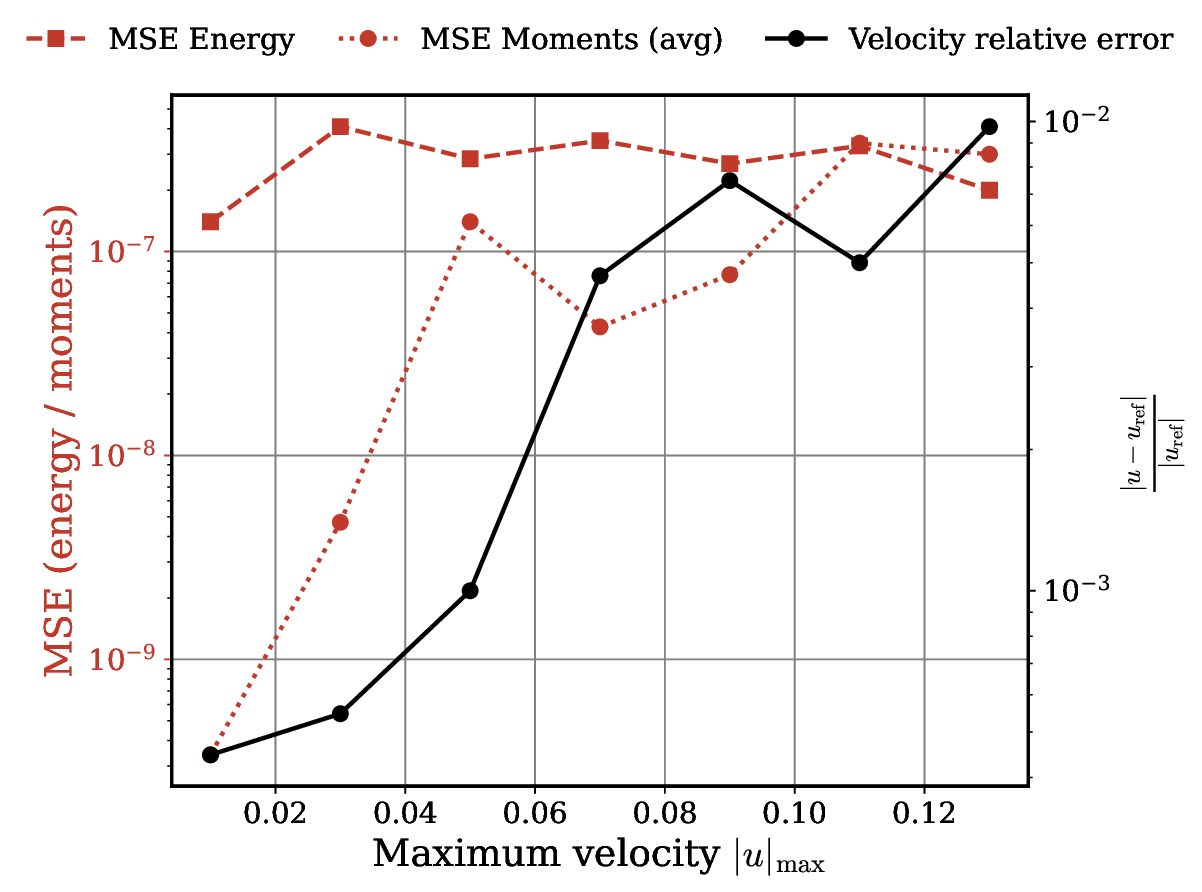}
\caption{Comparison of the mean square error (MSE) for energy $E$ and stress moments, and the relative error of the velocity, evaluated across different maximum velocities $|u|_{max}$ for the QML model. }
    \label{fig:energy_moments_vel}
\end{figure}

A closer look to Fig~\ref{fig:energy_moments_vel} shows that the momentum MSE using $p_{xx}-p_{yy}=f_1+f_3-f_2-f_4$ and $p_{xy}=f_5+f_7-f_6-f_8$, worsen for higher velocities as well. On the other hand, the total energy MSE with $E=-4f_0-f_1-f_2-f_3-f_4+2f_5+f_6+2f_7+2f_8$ is not impacted by the velocity. This result is expected as the flow is not turbulent and nearly incompressible. On the other hand, a perhaps more interesting observation is that the relative error of the velocity also increases at higher velocities. This is especially more pronounced at lower velocities, where the relative error increases rapidly. In contrast to our previous reasoning in Fig~\ref{fig:nonlinear_different_u}, where lower $\eta$ values are observed at higher velocities, this new evidence suggests that the model will perform better at low velocities in simulations. However, the results are still moderately optimistic for flows with velocities up to $u=0.1$. Notice that in the training dataset (extracted from a TGV simulation with 32x32 lattice sites), the velocities of each set of distribution functions are equally represented, with the same number of samples at each order of magnitude. However, internal tests show that the trend is identical across datasets with high velocity but a narrow, equivalent velocity range of $\Delta u=0.01$. 

\begin{figure}[htbp]
    \centering
        \includegraphics[width=0.48\textwidth]{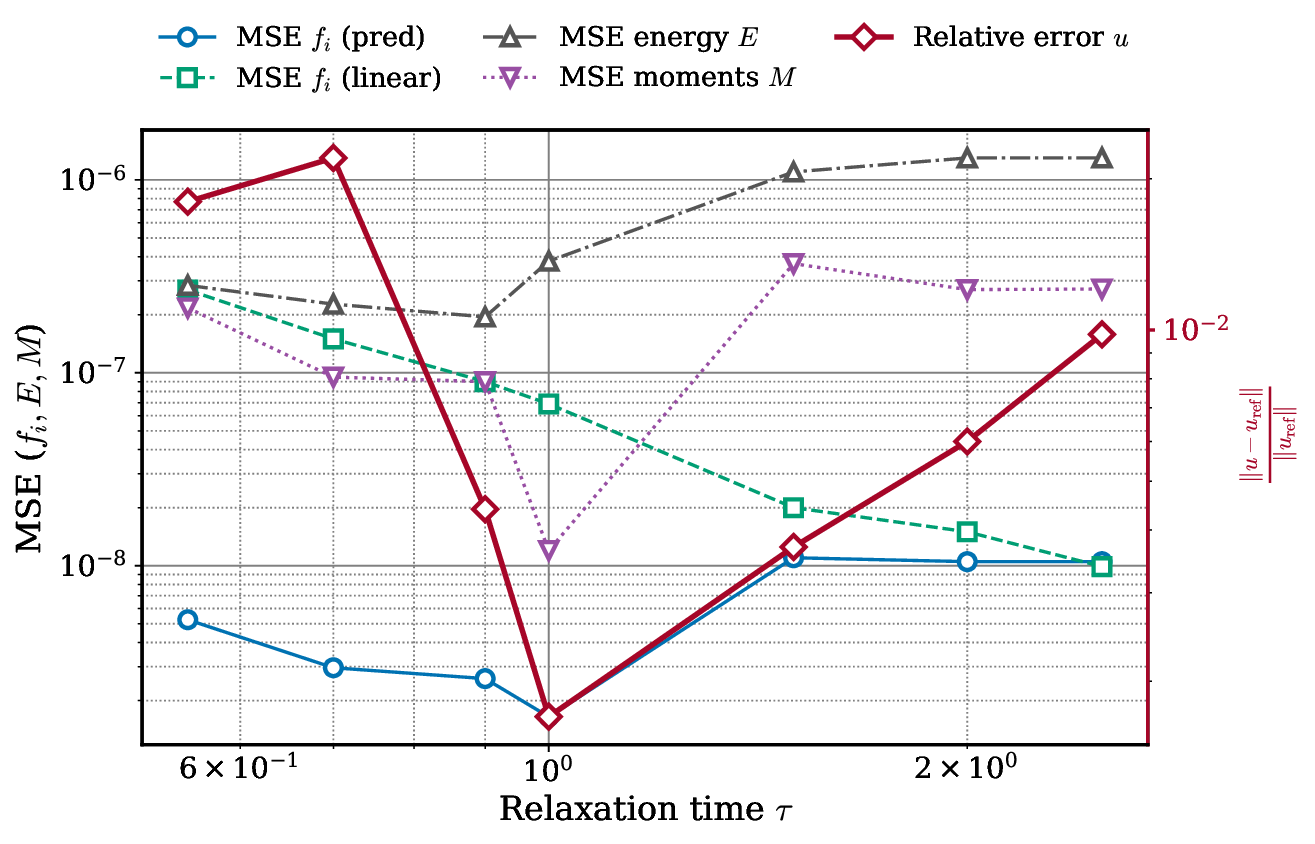}
    \caption{Comparison of the MSE of the predicted distribution function $f_i^{vqc}$ and post-linear collision distribution function $f_i^{lin}$ compared to $f_i^{ref}$, energy and momenta $[p_{xx}-p_{yy},p_{xy}]$ using $L=L_{A\phi}$ as loss function. The figure also includes the absolute relative error of the velocity on the right axis. The simulation is carried at $u=0.05$.}
    \label{fig:tau_comparison}
\end{figure}

Regarding the training at different viscosities, which is related to the relaxation time $\tau$ in LBM under the following expression,
\begin{equation}
    \nu=c_s^2(\tau-0.5)
\end{equation}
 we found that it affects the results as well. Our research confirmed the affirmation from \cite{lactatus2025surrogate}, where low accuracy is observed for $\tau$ values other than 1  when predicting the total LBM collision. However, contrary to the belief from previous research, this is not due to the MSE of the distribution function $f_i^{vqc}$, but rather to a large deviation from momentum conservation.  In Fig~\ref{fig:tau_comparison}, we can see that the MSE of momenta and $f_i^{vqc}$ and the absolute relative error for the velocity $u$ are best at $\tau=1$. However, the ratio between the MSE $f_i^{vqc}$ and the MSE of $f_i^{lin}$ compared to $f_i^{ref}$ (what we previously called $\eta$) is maintained or even improved at lower relaxation times. Despite that, the biggest source of error, the relative error of the velocity, quickly deteriorates. This indicates that, while not limited to $\tau=1$, the method behaves best using it. Larger values will have a negative effect on the prediction of stress momenta, velocity and distribution functions, while for the overrelaxation regime at $\tau<1$, the relative error of the velocity is very large.

\begin{figure}[htbp]
    \centering
        \includegraphics[width=0.48\textwidth]{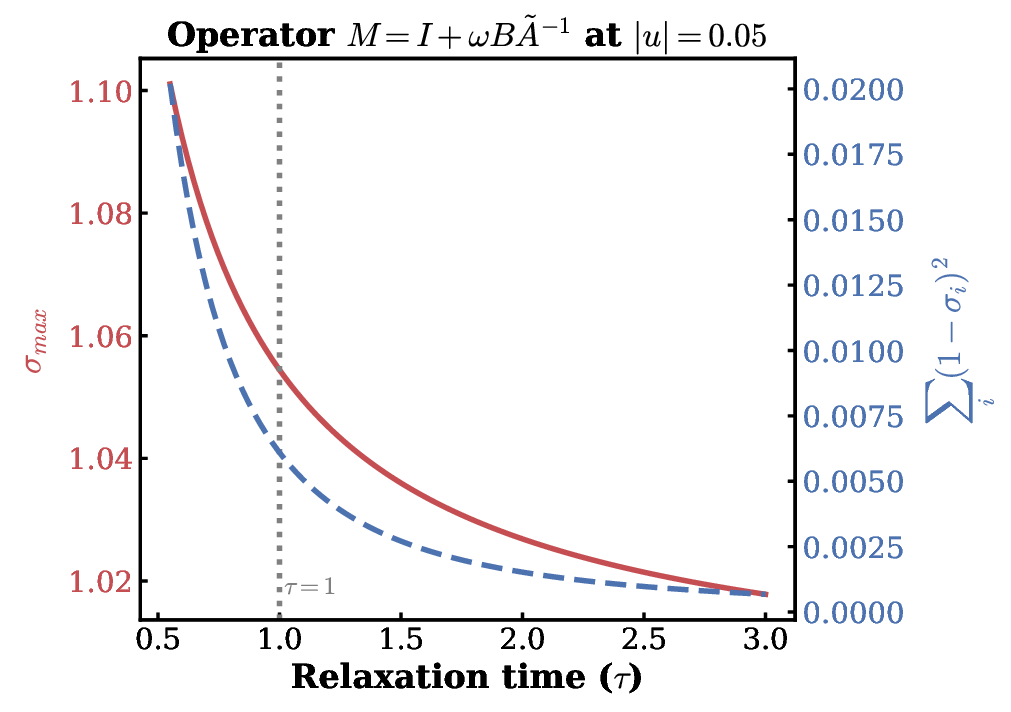}
    \caption{Nonunitary measure and singular value of $I+\frac{1}{\tau} B(u=0.05)\tilde{A}^{-1}$, which describes the linear transformation from $f_i^{lin}$ to $f_i^{ref}$}
    \label{fig:spread_tau}
\end{figure}

To identify the possible cause of this difference, we can take a look at the non-unitarity measure of the effective collision operator the model is learning. We will write the equations using $\omega=\frac{1}{\tau}$. The linear and nonlinear distribution functions ($f^{lin}$ and $f^{ref}$) are
\begin{equation}
\begin{split}
f^{lin} &= (1-\omega) f^{str} + \omega A f^{str} \\
f^{ref} &= (1-\omega) f^{str} + \omega \bigl(A f^{str} + B(u) f^{str}\bigr)
\end{split}
\end{equation}
which means that we can write
\begin{equation}
\begin{split}
f^{ref} 
    &= f^{lin} + \omega B f^{str} \\
    &= f^{lin} + \omega B \,(I + \omega(A-I))^{-1} f^{lin} \\
    &= \bigl(I + \omega B \tilde{A}^{-1}\bigr) f^{lin}
\end{split}
\end{equation}
Now, if we inspect the singular values of this operator and its non-unitarity measure, we obtain the Figure~\ref{fig:spread_tau}. It shows that larger values of $\tau$ correspond to more unitary operators, which agrees with the nonlinear contribution (a higher deviation from the linear and nonlinear distribution functions at lower $\tau$). This explains well the higher error we observe at lower $\tau$, but it does not clarify the differences observed at higher values. Contrary to previous research \cite{lactatus2025surrogate,Itani2025QMLLBM}, we do not think the cause lies in the lack of a copy of the original distribution function, which, in their case, is $f^{str}$. We will see this clearer in the next section (Sec~\ref{sec:R2}). A possible explanation for the observed phenomena is the presence of non-equilibrium components in the non-conserved moments at $\tau > 1$. At higher values of $\tau$, these non-equilibrium components of $f_i$ generate ghost moments and other non-conserved contributions, which can make predicting the conserved moments (in this case, the velocity) more difficult. This also explains why the mean squared error of $f^{\rm vqc}$ and the non-conserved moments $M$ remain essentially unchanged for $\tau > 1$.

Finally, we tested whether the restriction that enables coherent simulations without high error rates, maintaining all phases at zero, introduces any problems. To test this, we compared the maximum relative error in the velocity field for a TGV simulation at $u=0.05$, $T=50$, and $\tau=1$ when measuring at each time step (to isolate the amplitudes) to determine whether we lose accuracy by forcing the phase predictions. The results can be seen in Table~\ref{tab:amplitude_amplitudephase}, where no substantial difference can be observed. The model's trainability variability is most likely responsible for the observed difference rather than a fundamental change. We conclude that targeting the phases and the amplitudes presents no disadvantage compared to training the model only with the amplitudes. However, targeting the phases allows us to realise the simulation without intermediary measurements, as the operator is unitary, linear and we reconstruct the quantum state at each time-step. Additionally, we do not observe any increment in the training time when adding an accurate phase compared with the usage of $L_A$ (loss function without the phase contribution). Finally, we want to make a remark on $\eta_\epsilon=\frac{\epsilon_u^{lin}}{\epsilon_u^{QML}}$ (ratio between the linear $\epsilon_u^{lin}$ and QML $\epsilon_u^{QML}$  relative error at the end of the simulation). As shown in the table, the improvement achieved by the QML over the linear case decreases at higher velocities.

\begin{table}[H]
\centering
\begin{tabular}{|c|c|c|c|}
\hline
$u_{max}$ & Max relative error-$L_{A\phi}$ & Max relative error-$L_{A}$ &$\eta_\epsilon$\\
\hline
0.01   & 0.0099  & 0.01&4.95 \\
0.03  & 0.036  & 0.035&4.22 \\
0.05  & 0.092 & 0.093& 2.89\\
0.075  & 0.23 & 0.24&1.77  \\
0.1 & 0.48 & 0.49& 1.16 \\

\hline
\end{tabular}
\caption{Comparison between the maximum relative error when targeting the amplitudes and both the amplitudes and the phases at different velocities. The error rate shown for $L_{A}$ assumes measurement at each time-step, while for $L{A\phi}$ this is not the case. The value $\eta_\epsilon=\frac{\epsilon_u^{lin}}{\epsilon_u^{QML}}$ with $\epsilon_u^{lin}$ the maximum relative error in the simulation comparing GBK LBM at second order with the linear one and $\epsilon_u^{QML}$ compares the GBK LBM simulation with the QML result.}
\label{tab:amplitude_amplitudephase}
\end{table}

\subsubsection{Relaxing the unitarity condition}
\label{subsec:R1_3}

Following the evidence seen in Fig~\ref{fig:spread}, we have focused on studying the behaviour and best practice of a VQC based on one important premise: The unitarity of the operator. Ideally, we would like the nonlinear collision operator to be unitary at every time step. This means that the simulation must be coherent. However, in this chapter, we will explore whether relaxing this condition offers advantages in terms of accuracy that we can trade off against by increasing the model's non-unitarity. To do this, there exist two main approaches. The first is to use a similar approach to the linear combination of unitaries (LCU) \cite{childs2012hamiltonian}, where we use additional ancilla to increase the Hilbert space and store our desired results in a subspace that we will later measure. However, this technique has a major disadvantage. By including an additional ancilla, it becomes hard to find the ansatz that best preserves the equivariances from LBM while maintaining flexibility. Therefore, we will allow the amplitudes of the states not encoding any information (in this case $f_9$ to $f_{15}$) to have a non-zero amplitude, which later can be distinguished with an ancilla qubit and measured at each time-step as done for LCU. 

\begin{figure}[htbp]
    \centering
        \includegraphics[width=0.48\textwidth]{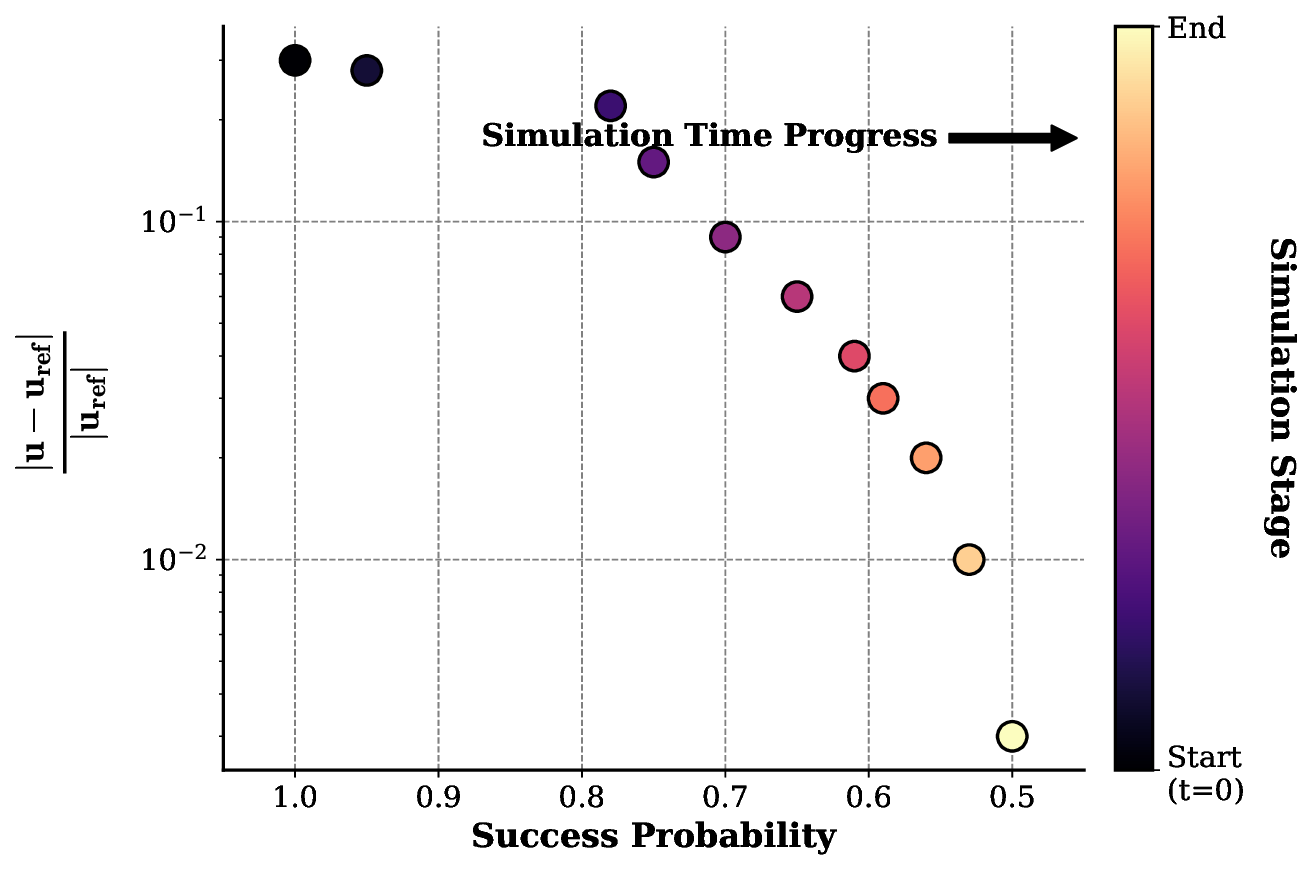}
    \caption{Trade-off between the maximum relative error in the velocity field and the total probability to measure one of the first 9 states corresponding to physical $f_i$ in the D2Q9 LBM scheme when the non-linear collision is trained with a variational quantum circuit. A TGV flow is used with $u_{max}=0.05$, $T=50$, $\tau=1$. The loss-function used to train the model is the MSE of the amplitude of the distribution function between $|f^{vqc}|$ and $|f^{ref}|$. During the simulation a measurement is taken at each time-step. The colour scale represents the passage of time from the beginning to the end of the training.}
    \label{fig:tradeoff}
\end{figure}

First, we will focus on studying if the loss of unitarity, where we use $\frac{f_i^{vqc}}{\sum\limits_{i=1}^9 f_i^{vqc}}$ as the target distribution function, provides a lower maximum relative error in the velocity field. In Figure~\ref{fig:tradeoff}, we can see how the relative error of the velocity decreases during the training of the model while the success probability decreases from 1 to 0.5. Notice that the relative error here is not the mean relative error of the training dataset but the one obtained from a real simulation using the learned unitary for each data point. Additionally, we observed that the model was faster and more consistent during training compared to the unitary case. This demonstrates the model's ability to approximate nonlinear terms under these conditions with very high accuracy. Compared to the model targeting $|f^{ref}|$ without normalisation in the first 9 amplitudes and unitary, we obtain 50 times lower error. Notice that in this case we are not considering the phases and measuring at each time-step.

When we consider the phases, we observe a similar level of accuracy, with the model becoming more sensitive to the phase than its unitary counterpart, which increases the training difficulty and the consistency of the results the model provides. Furthermore, the success probability increases for this case and the model leaks less amplitude to undesired states without imposing it in the cost function. Additionally, the phases were sensitive to the dataset used. Using a TGV simulation at a lower resolution (32x32 lattice sites) and data from a turbulent Kolmogorov flow, the MSE in the phase was of orders of magnitude. Our interpretation is that the model leverages interference across the full 16-amplitude space to achieve higher accuracy, resulting in a more sensitive dependence. Additionally, we observed that the distribution of the phase MSEs matched the non-equilibrium contributions in the training dataset. This suggests that the model is unable to handle the nonlinearities, even in the nonunitary case, with the ansatz used. 

Another important consideration is that when the phases are added for the nonunitary case, the tradeoff seen in Figure~\ref{fig:tradeoff} is not observed. The final success probability remains relatively high, and the phase accuracy limits the model's final accuracy. This behaviour is indicative of the model's sensitivity to the initial phase during the training. We tested training the model by sampling random noise in the initial phase at each time step to increase its resilience, but no improvements were observed. Nonetheless, we still believe future studies should tackle this possibility. 

Despite these issues, we obtained higher accuracy than the unitary case considering both amplitudes and phases (Table~\ref{tab:amplitude_amplitudephase_nonunitary}) and therefore higher $\eta_\epsilon$ for every velocity range. We conclude that the best practice is to use the fully unitary model for combining multiple time-steps, as it is also easier to train. Future research must address whether it is possible to reduce the phase's error for the nonunitary case, which remains the limiting factor here. 

\begin{table}[H]
\centering
\begin{tabular}{|c|c|c|c|}
\hline
$u_{max}$ & Max relative error with $L_{A\phi}$ &$\eta_\epsilon$\\
\hline
0.01   & 0.0065  &8.33 \\
0.03  & 0.021  & 6 \\
0.05  & 0.065 &3.85 \\
0.075  & 0.17 & 2.42 \\
0.1 & 0.39 & 1.46  \\

\hline
\end{tabular}
\caption{Comparison between the maximum relative error for the nonunitary case when targeting the amplitudes and both the amplitudes and the phases at different velocities. The error rate showon for $L_{A}$ assumes measurement at each time-step, while for $L{A\phi}$ this is not the case.}
\label{tab:amplitude_amplitudephase_nonunitary}
\end{table}

\subsubsection{Nonlinear test cases}
\label{subsec:R1_4}
We first reviewed the conditions under which the R1 model is most efficient. Since the nonunitary model offers no advantage for learning both amplitudes and phases in simulations with more than one time step and no intermediate measurements, we now evaluate the R1 model on several test cases. Previous research \cite{lactatus2025surrogate,Itani2025QMLLBM}, is mainly focused on showing the efficiency of a learned unitary collision operator at low Reynolds number without special focus to nonlinear flows. While famous test cases known to be highly nonlinear are shown, like Taylor Green Vortex or lid-driven cavity, these are only simulated at low Reynolds number, where the difference between the LBM collision operator using only linear terms and adding $O(u^2)$ terms is much lower than the error rate observed. A clear comparison is not displayed either. These tests are inadequate to demonstrate the model's ability to handle nonlinear terms. Additionally, the simulations are performed without accounting for phase errors, and a measurement is used at each time step.

In order to put in evidence if we can obtain an advantage in the nonlinear realm, we will show the evolution of the relative error rates between our model and the target simulation and between the linear and the target. We will focus on three cases beyond the TGV used so far. First, a Kolmogorov transient flow where an initial force is added, and we want to predict the velocity field and the decay of the maximum velocity over time. Second, a flow with an object (flat plate) where we focus on the error of the vorticity field created by the nonlinear terms. Third, we consider 2D orthogonal crossing jets, where a force is added throughout the domain at each time step, and compare the relative error in the velocity field. These three examples cover the transient, steady, and forcing flows, which we believe provide a clear view of this model's capabilities. To avoid questioning the generality of the training dataset and to simplify the analysis, we will use a set of $2\cdot 10^5$ data points extracted from the real simulation. This means that the results here will most likely present the best-case scenario. 

In the first simulation case, we chose a Kolmogorov-like flow with a grid size $N=L_x L_y$ where an initial velocity field is selected, and its maximum velocity decreases over time. As an initial condition, the distribution functions are initialised as 

\begin{equation}
\begin{aligned}
f_i(x,y)=w_i\Big(1
&+A_x\cos\!\left(\frac{2\pi}{L_y}k_x y\right)c_i\cdot c_1 \\
&+A_y\cos\!\left(\frac{2\pi}{L_y}k_y x\right)c_i\cdot c_2
\Big)
\end{aligned}
\label{eq:kolmogorov_initial}
\end{equation}
 with $k_x$,$k_y$ the wave numbers, $A_x$,$A_y$ positive amplitudes and $c_i$ the velocity direction of each $f_i$.

\begin{figure}[htbp]
    \centering
    
    \begin{subfigure}[t]{0.48\textwidth}
        \centering
        \includegraphics[width=\textwidth]{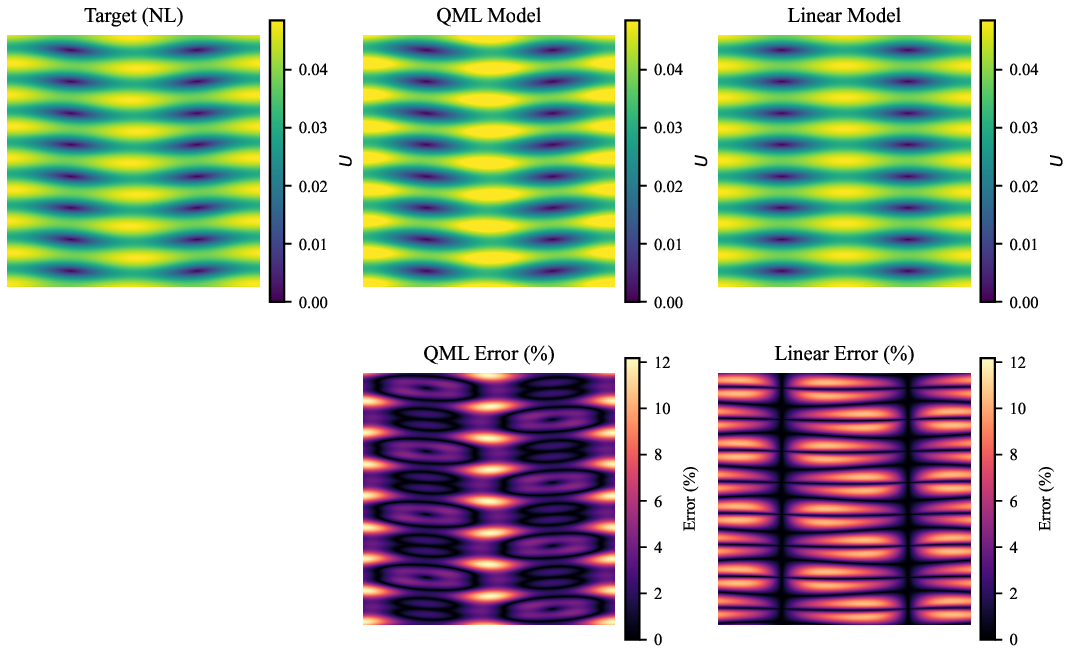}
        \caption{2D velocity field comparing the target nonlinear LBM collision, the QML model and the linear collision LBM. The absolute error between the velocity field of the target and QML and the target and linear models are also displayed.}
    \end{subfigure}
    \hfill
    \begin{subfigure}[t]{0.48\textwidth}
        \centering
        \includegraphics[width=\textwidth]{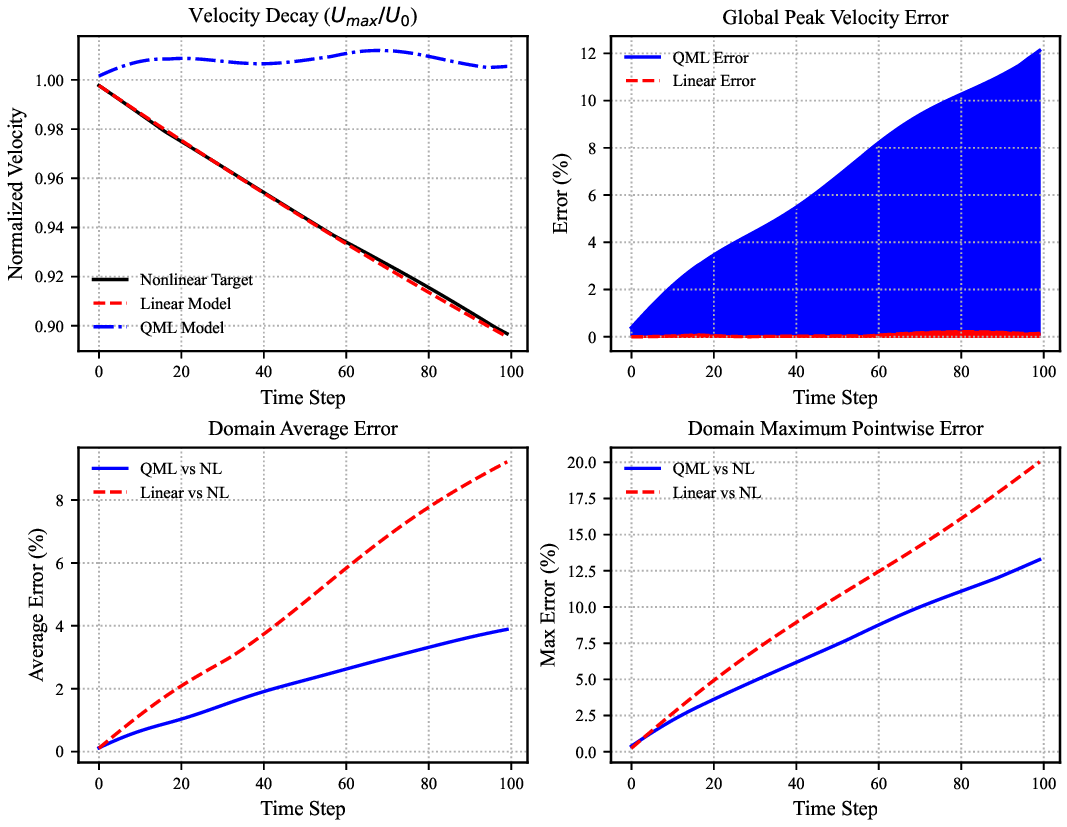}
        \caption{The velocity decay expressed as $u_{max}/u_0$ with $u_{max}$ the maximum velocity at each time-step and $u_0$ the maximum velocity at the beginning of the simulation. In the upper-right, we show the absolute error between the QML velocity decay and the nonlinear and the linear one over time. Finally, at the bottom, the average and maximum pointwise absolute errors from the simulation are displayed.}
    \end{subfigure}

    \caption{The maximum velocity decay of a Kolmogorov flow is simulated by imposing an initial velocity field with $A_x=0.18$ and $A_y=0.09$, $k_x=1$, $k_y=1$ following \eqref{eq:kolmogorov_initial}, $\tau=1$, 100 time-steps and $L_x=L_y=256$ square domain.}
    \label{fig:kolmogorov_relaxation}
\end{figure}

The goal of this simulation is to demonstrate the model's inability to predict the velocity field with high accuracy. In this case, as the decay of the maximum velocity happens quickly and linearly over time, we can observe its accuracy well. Additionally, no objects or forces disturb the flow, meaning no collisions occur outside the quantum computer, such as with walls or added forces. Figure~\ref{fig:kolmogorov_relaxation} contains the results of the simulation. While the QML was unable to predict the evolution of the velocity decay, its average and pointwise errors were lower in absolute terms. A similar conclusion is reached when calculating the relative error. Additionally, as shown in the figures, the target (NL) and the QML models exhibit curvature in the velocity profile due to the nonlinear part of the collision, which is not observed in the linear 2D velocity field. We conclude that while the QML cannot achieve high accuracy in the velocity field, it still exhibits behaviour characteristic of the nonlinear part, demonstrating the ability to simulate nonlinear flows on the quantum computer. Notice that in this case both amplitudes and phases are predicted at each time-step.

\begin{figure}[htbp]
    \centering
        \includegraphics[width=0.48\textwidth]{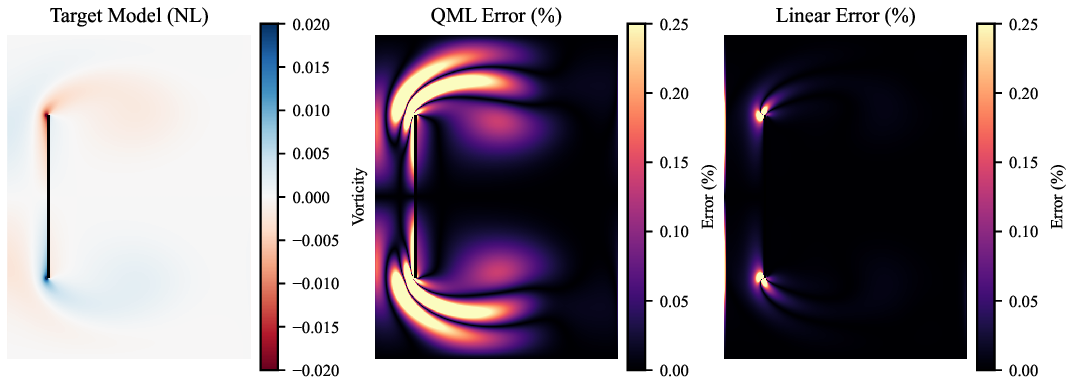}
    \caption{LBM simulation using the GBK operator with an inlet velocity $u=0.025$ and a flat plate as obstacle. The simulation is carried out over 10000 time-steps, during which the last 10 time-steps are computed using the QML and the linear operators in parallel. The error 2D plots show the absolute error between the target nonlinear model (NL) and the approximated QML and linear flows. A grid of $L_x=400$ and $L_y=600$ and $Re=35$ with $u_{max}=0.025$ is used.}
    \label{fig:flat_plate}
\end{figure}

Next, for our second test case, we use a flat plate perpendicular to the flow direction (positive x-direction). We choose a flat plate due to its high drag coefficient, $C_d=1.28$, which leads to high vorticity at low Reynolds numbers. In this case, we want to estimate if the QML model is able to preserve the vorticity field once formed for at least few time-steps. Collisions with the object are always treated classically. Because vorticity depends on the velocity gradient and is caused by the nonlinear part of the collision operator, we initially expected low accuracy. In Figure~\ref{fig:flat_plate} we can corroborate our initial hypothesis. In this case, we simulated $T=10000$ time-steps in a domain with $L_x=400$ and $L_y=600$ and $Re=35$ with $u_{max}=0.025$. The simulation is carried out for $T=10000$, but only the last 10 time-steps are realised using the linear and QML models. The results clearly show how the QML is incapable of maintaining the vorticity field, while the linear model is very close to it.  As expected, the QML produces results with higher error than the linear case. This is caused by the fictitious force acting added at each time-step by the QML model as a result of a limited precision. For simulations with high vorticity, where this is important, the QML is not ideal, and further research is needed to ensure the collision operator conserves momentum at each time step.

\begin{figure}[htbp]
    \centering
    
    \begin{subfigure}[t]{0.48\textwidth}
        \centering
        \includegraphics[width=\textwidth]{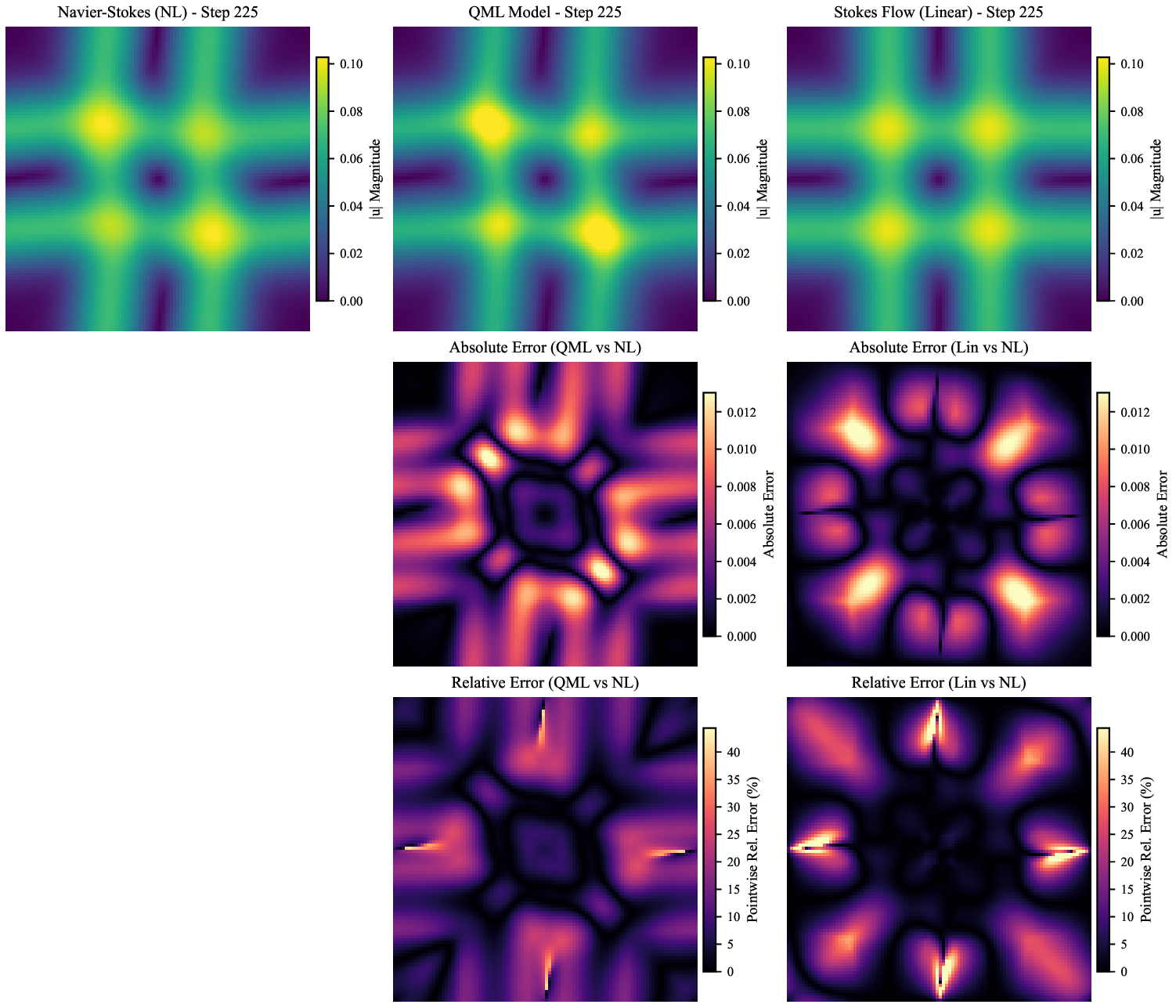}
        \caption{2D velocity fields comparison between the nonlinear, QML-based and linear collisions. For QML and linear collisions, the absolute and relative errors between these and the nonlinear final 2D velocity field are shown.}
    \end{subfigure}
    \hfill
    \begin{subfigure}[t]{0.48\textwidth}
        \centering
        \includegraphics[width=\textwidth]{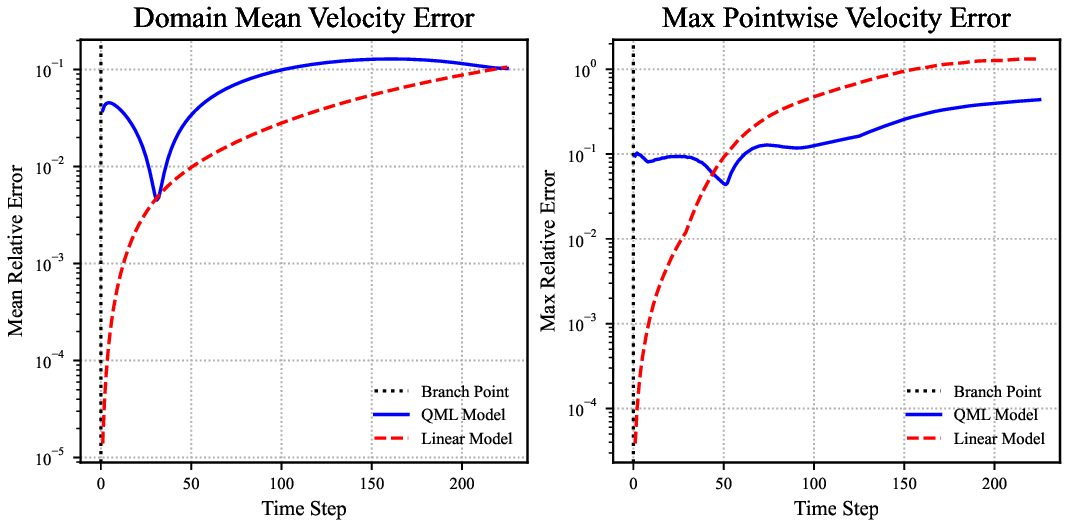}
        \caption{Average and pointwise relative error in the domain along time when comparing the QML and linear models with the nonlinear one.}
    \end{subfigure}

    \caption{Comparison between the nonlinear collision, QML-based collision and linear collision for counter-propagating Gaussian jets. The simulation parameters are:  $\tau=1$, $G_0=4\cdot 10^{-4}$, $W=10$, $y_{h_1}=\frac{N_y}{3}$, $y_{h_2}=\frac{2N_y}{3}$, $x_{v_1}=\frac{N_x}{3}$, and $x_{v_2}=\frac{2N_x}{3}$ on a domain of size $N_x \times N_y = 100 \times 100$ for $T=225$ time-steps. For the QML model, the force added each time step is $20\%$ larger to counter the fictitious force added at each time-step.}
    \label{fig:jet_1}
\end{figure}

Finally, for our third case, we consider the interaction of orthogonal, counter-propagating Gaussian jets. To isolate the fluid interactions and avoid wall collisions, we use periodic boundary conditions. In this simulation, we want to explore how the model behaves when a force is applied across the entire domain at each time step. Given that the QML model introduces fictitious forces at each time-step, we expect this case to work particularly well when compared with the nonlinear case, as the relative error of the force introduced will be smaller. Specifically, we use a 2D force with Gaussian profiles given by

\begin{equation}
\begin{aligned}
G_x(y) &= G_0\left(\exp\!\left(-\frac{(y-y_{h_1})^2}{W^2}\right)
           - \exp\!\left(-\frac{(y-y_{h_2})^2}{W^2}\right)\right) \\
G_y(x) &= G_0\left(\exp\!\left(-\frac{(x-x_{v_1})^2}{W^2}\right)
           - \exp\!\left(-\frac{(x-x_{v_2})^2}{W^2}\right)\right)
\end{aligned}
\end{equation}
with $G_0$ the force amplitude, $y_{h_1}$, $y_{h_2}$, $x_{v_1}$ and $x_{v_2}$ the domain's fractional coordinates and $W$ the width of the jets. In Figure~\ref{fig:jet_1}, results are shown for a simulation with $\tau=1$, $G_0=4\cdot 10^{-4}$, $W=10$, $y_{h_1}=\frac{N_y}{3}$, $y_{h_2}=\frac{2N_y}{3}$, $x_{v_1}=\frac{N_x}{3}$, and $x_{v_2}=\frac{2N_x}{3}$ on a domain of size $N_x \times N_y = 100 \times 100$ for $T=225$ time-steps. In the QML case, the force added at each time-step is $20\%$ larger to counteract the fictitious force, which reduces the flow velocity in this example.

\begin{figure}[htbp]
    \centering
    
    \begin{subfigure}[t]{0.3\textwidth}
        \centering
        \includegraphics[width=\textwidth]{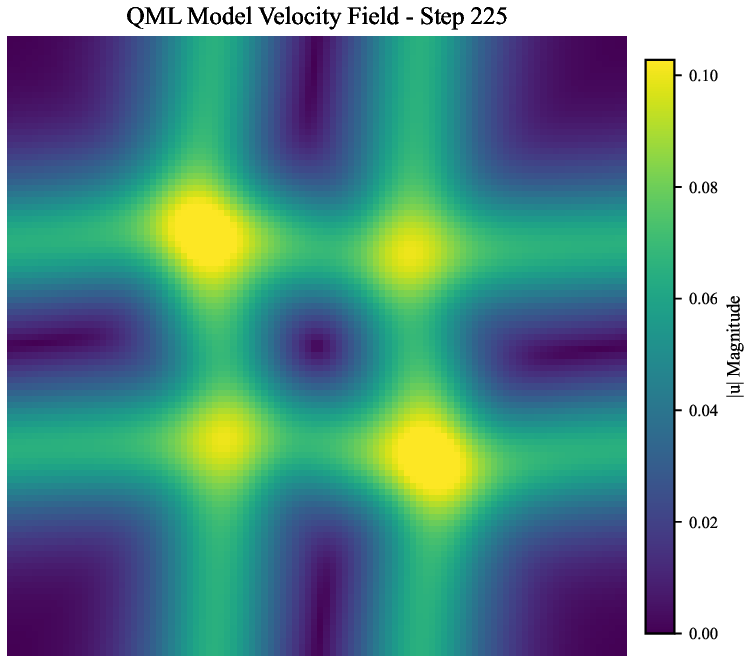}
        \caption{Velocity field with a gaussian force with $G_0=5.28\cdot 10^{-4}$ added at each time-step.}
    \end{subfigure}
    \hfill
    \begin{subfigure}[t]{0.3\textwidth}
        \centering
        \includegraphics[width=\textwidth]{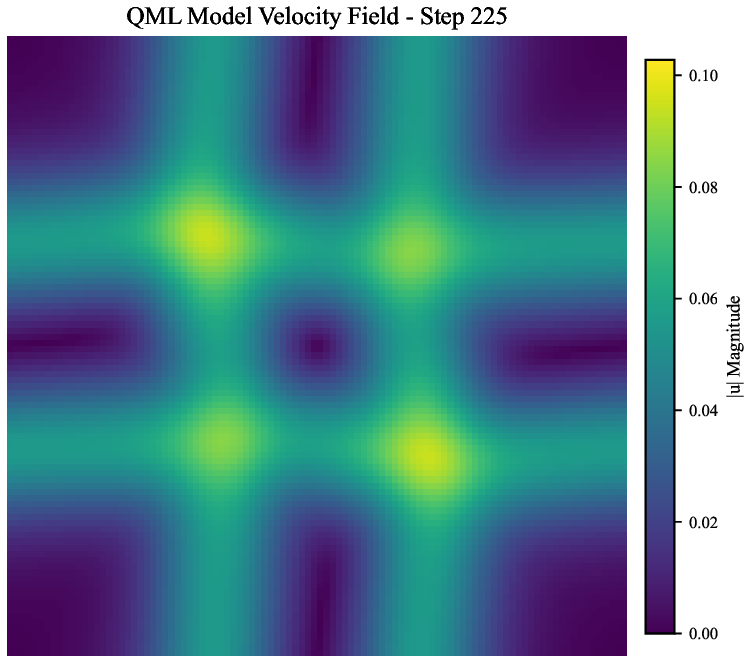}
        \caption{Velocity field with a gaussian force with $G_0=4.4\cdot 10^{-4}$ added at each time-step.}
    \end{subfigure}

    \caption{Comparison between the 2D velocity fields of two 2D orthogonal crossing jets simulations using the QML model with different forces added at each time-step}
    \label{fig:jets_different_force_added}
\end{figure}

The results show a higher average relative error of the QML model velocity flow during the simulation, while the maximum pointwise relative error is lower. However, when we examine the final velocity profiles, we see that the curvature obtained by the QML model is similar to that of the nonlinear simulation. In fact, the relative and absolute error plots show an incorrect velocity in the jets but a similar shape. If we do not adjust the force and use the same value for the linear, QML, and nonlinear simulations, we obtain a higher relative error. This comparison can be observed in Figure~\ref{fig:jets_different_force_added}, where a similar shape is obtained, but the velocity is damped. We attribute the damping to the fictitious forces added by the QML model. The strategy shown here of adding force to account for reduced accuracy in momentum conservation is presented solely for test purposes. In general, it will be hard to identify which parameter can be modified to obtain higher accuracy beforehand. After all, the QML is not truly adding fictitious forces beyond our comprehension, but simply adding them as a result of a limited number of significant figures predicted (both in the velocity used to calculate the nonlinear terms and in the nonlinear contribution created with that velocity).

\begin{figure*}[htbp]
    \centering
    \includegraphics[width=1\textwidth]{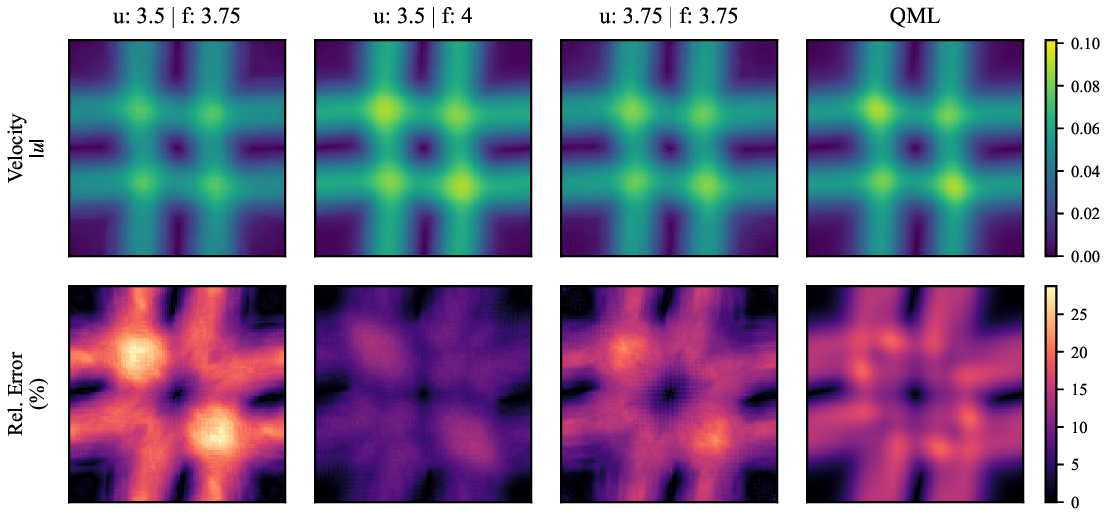}
    \caption{Comparison between the 2D velocity fields of two 2D orthogonal crossing jets and their relative errors. The numbers next to $u$ and $f$ indicate the decimal precision of the registers used for the velocity $u$ and for the nonlinear part and final distribution functions in the operator $f$. Decimal values are calculated as a weighted average of the floor and ceiling integers surrounding the given value.}
    \label{fig:significant_figures}
\end{figure*}

Considering the average MSE error for this particular case during training is $10^{-8}$, roughly four decimal places are correctly estimated (which is reduced for higher $Re$ when the difference between $f^{lin}$ and $f^{ref}$ increases). Figure~\ref{fig:significant_figures} shows how this particular simulation differs when considering a fixed decimal precision when calculating the velocity and the equilibrium distribution function. Decimal precision between two integers as 3.5 is calculated as a weighted average of the floor and ceiling integers surrounding the given value. According to previous research \cite{Lehmann2022}, at least six decimal places of precision are usually needed in LBM to obtain accurate results, with no substantial difference observed between six and twelve decimals. In their article, the authors show that four decimals to store variables during the simulation can be sufficient if a specific encoding and methodology proposed by the same authors is followed. Despite that, six decimals are required during the computation. We believe that future research can involve exploring this or other encodings where lower precision is needed to obtain accurate results. 

\begin{figure}[htbp]
    \centering
    
    \begin{subfigure}[t]{0.48\textwidth}
        \centering
        \includegraphics[width=\textwidth]{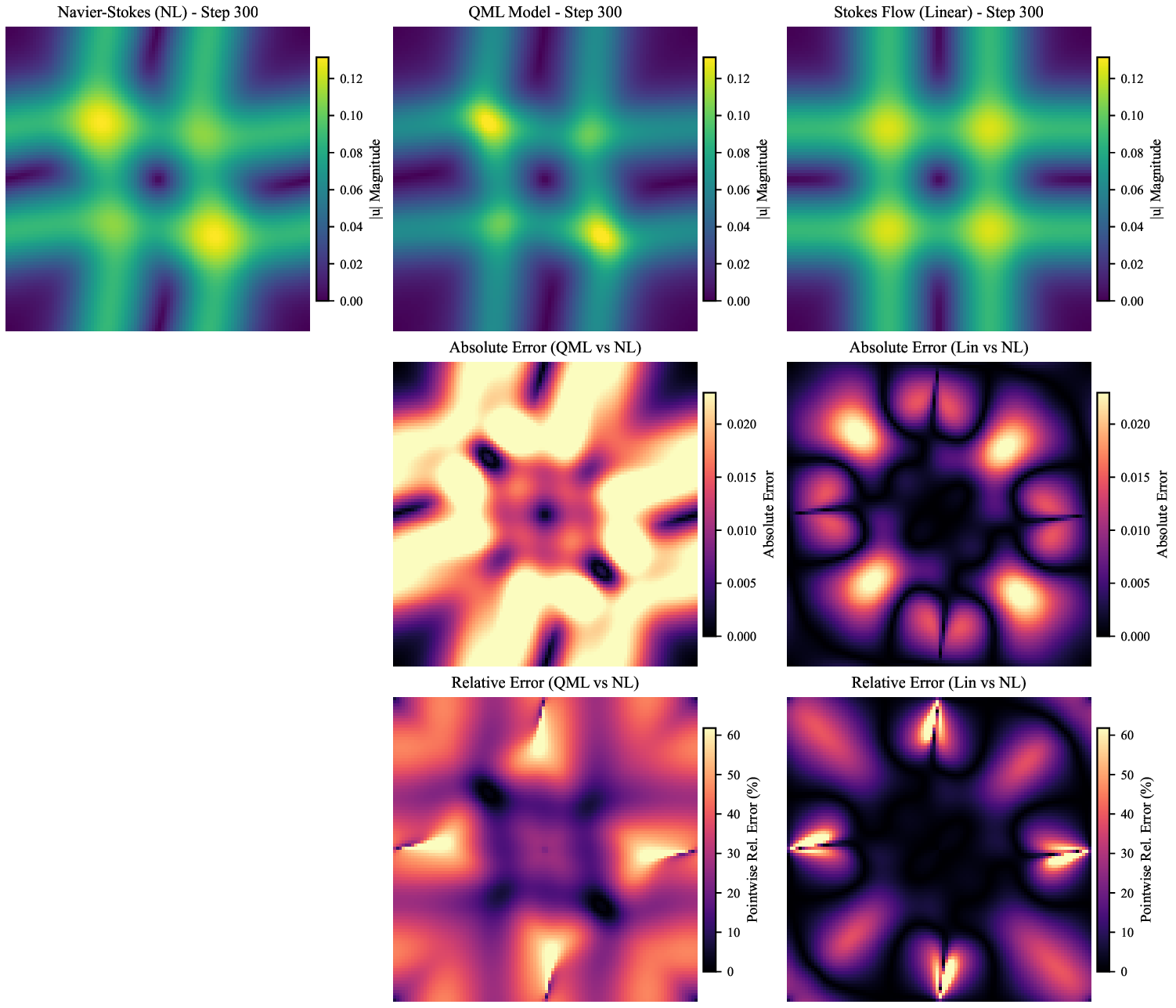}
        \caption{2D velocity fields comparison between the nonlinear, QML-based and linear collisions. For QML and linear collisions, the absolute and relative errors between these and the nonlinear final 2D velocity field are shown.}
    \end{subfigure}
    \hfill
    \begin{subfigure}[t]{0.48\textwidth}
        \centering
        \includegraphics[width=\textwidth]{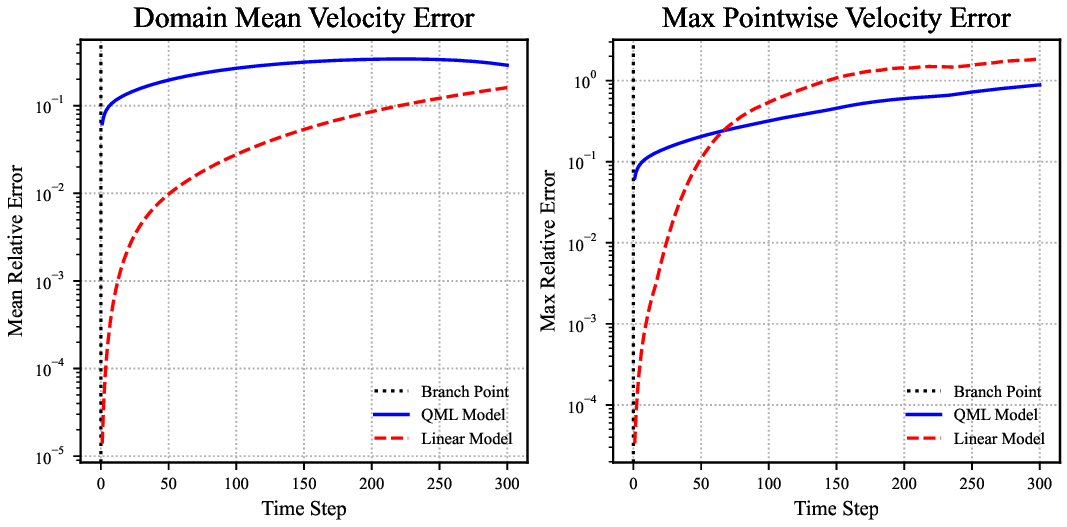}
        \caption{Average and pointwise relative error in the domain along time when comparing the QML and linear models with the nonlinear one.}
    \end{subfigure}

    \caption{Comparison between the nonlinear collision, QML-based collision and linear collision for counter-propagating Gaussian jets. The simulation parameters are:  $\tau=1$, $G_0=4\cdot 10^{-4}$, $W=10$, $y_{h_1}=\frac{N_y}{3}$, $y_{h_2}=\frac{2N_y}{3}$, $x_{v_1}=\frac{N_x}{3}$, and $x_{v_2}=\frac{2N_x}{3}$ on a domain of size $N_x \times N_y = 100 \times 100$ for $T=300$ time-steps. For the QML model, the force added each time step is $25\%$ lower to counter the fictitious force added at each time-step.}
    \label{fig:gaussian_jets_300}
\end{figure}

Finally, we went beyond the training of the model, and continued the simulation up to $T=300$ to make the effects more visible. As the QML model was trained with velocities $u<0.1$, we reduced the force in this case instead of increasing it for this particular model (not for the linear and nonlinear simulations) to $3\cdot 10^{-4}$. Figure~\ref{fig:gaussian_jets_300} show a very high error for the velocity field of the QML compared to the Navier-Stokes. Despite that, the velocity field is shaped similarly. Most likely, this effect is produced by the highly accurate prediction of the nonlinear distribution functions $f_i^{vqc}$ and their stress tensors $p_{xy}$ and $p_{xx}-p_{yy}$. 

After these three simulations and our extensive R1 model analysis, we can conclude that the QML unitary is capable of modelling the nonlinear flows, at least in some simulation cases. In general, the produced unitary works particularly well for simulations where we target the shape of the flow rather than its specific velocity or where high forces are produced at each time-step. Consider that the Gaussian jets simulation includes a force in the domain, but still it is very small compared with the accuracy of the velocity at each time-step, which is of the order of $10^{-4}$. In other cases, such as the velocity decay in a Kolmogorov flow or the TGV, higher accuracy can be obtained in the average domain than in the linear case. Further research is needed on the conservation of momentum in the collision operator for the QML unitary. Other possible research directions include learning phases for the nonunitary case, where probability can leak to undesired states. Additionally, as introduced in \ref{sec:appendix}, some cases may benefit from the diffusive encoding. While in classical LBM this encoding is used to achieve higher accuracy at the expense of higher computational cost, in our case, we can benefit from the much lower MSE in the velocity field and higher accuracy overall at lower velocities. In some cases, reducing the velocity even with a larger number of time steps may result in a lower absolute error at the end of the simulation. Our tests indicate that this effect is most pronounced for simulations with a small number of time steps and high velocities. Further research and testing will be required to draw conclusive results, but this approach represents a promising direction for future investigation.

\subsection{VQC for QLBM with two registers: R2 model}
\label{sec:R2}
In this section, we will consider a more complex model to achieve higher accuracy and address the R1 model's biggest issue: the lack of momentum conservation in the collision operator. The problem is that the model lacks information about the flow's current velocity. Furthermore, as we try to solve the nonlinear part, the model fails to do so across a wide range of velocities. Not having a copy of the current distribution functions $f^{lin}_i$ in the system makes it very hard for the model to achieve high accuracy. To solve this issue, we will now use a model with two quantum registers, each encoding the same distribution function, and initialise them in a tensor product state. 
\begin{figure*}[htbp]
    \centering
    \includegraphics[width=1\textwidth]{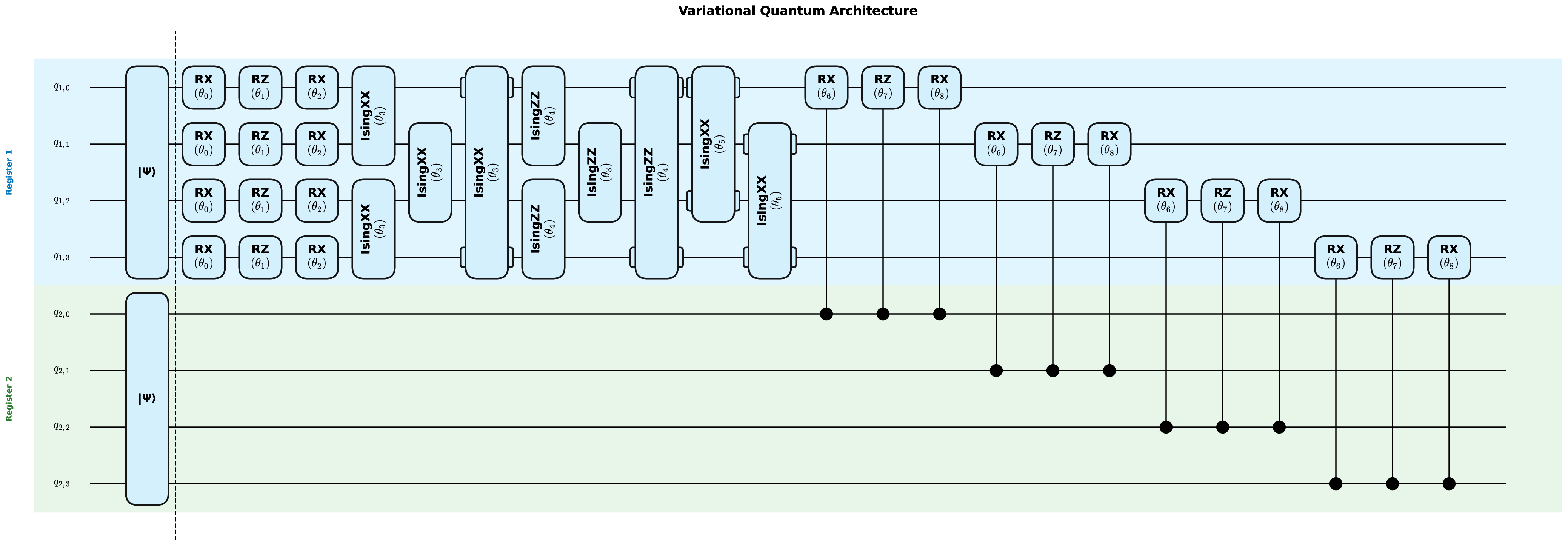}
    \caption{Schematic representation of the two-register Variational Quantum Circuit (R2 model) for the Quantum Lattice Boltzmann Method with one layer $B=1$. The architecture follows the same principles of previous research \cite{lactatus2025surrogate} to conserve dihedral symmetry as introduced in Sec~\ref{sec:dihedral_d2q9}. It also includes three different gates for each qubit, following the Euler angles principle for general rotations. Beyond that, two registers are used, such that the second register can inform the first one of the current relative amplitudes of each $f_i$.}
    \label{fig:2r_circuit}
\end{figure*}
In Fig~\ref{fig:2r_circuit}, we displayed the quantum circuit used using a single layer $B=1$ (we used $B=20$ to train the model; beyond that, the accuracy did not increase significantly). The goal of the second register is only to act as an informant to the first register. As no gate acts in the second register, the only change on the reduced density matrix $\rho_2$ of the second register will be a change of phase driven by the phase kickback effect. Notice that as we establish a tensor product of two registers and apply CNOT gates between them, the resulting state will be an entangled state. This means that $\rho_T\neq\rho_1\otimes\rho_2$. Training the quantum circuit to provide a closely similar result to $\rho^{ref}=\rho_1^{ref}\otimes\rho_2^{ref}$ is not possible, while targeting all the amplitudes is inefficient and slow. For this reason, we will use a cost function $L_{\rho_1}$ with $\rho_1$ the reduced density matrix of the first register, 
\begin{equation}
\begin{aligned}
L_{\rho_1} = \| \rho_1^{\mathrm{ref}} - \rho_1^{\mathrm{vqc}} \|_F^2.
\label{eq:loss_rho_2}
\end{aligned}
\end{equation}
As the quantum state evolves due to entanglement and training on reduced density operators, multi–time-step simulations without measurement are not considered in this model. Following our procedure with the R1 model, we will train it using a 2D TGV flow with $L_x=L_y=64$, $\tau=1$ and $u=0.1$. The training dataset will be generated from a TGV simulation with $L_x=L_y=32$ (we are not evaluating the model's generalisation here but its potential capability). 
\begin{figure*}[htbp]
    \centering
    
    \begin{subfigure}[t]{0.6\textwidth}
        \centering
        \includegraphics[width=\textwidth]{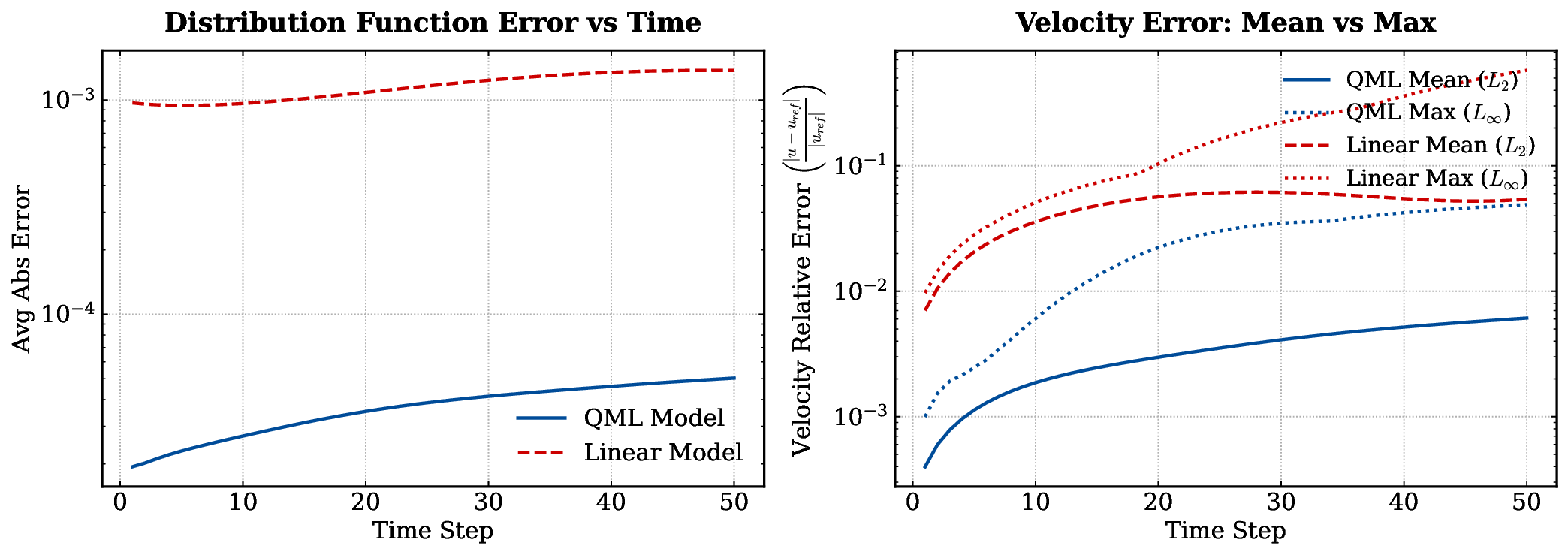}
        \caption{Evolution of the absolute error of $|f_i^{vqc}|$ and relative error of the velocity $u$ along time.}
    \end{subfigure}
    \hfill
    \begin{subfigure}[t]{0.6\textwidth}
        \centering
        \includegraphics[width=\textwidth]{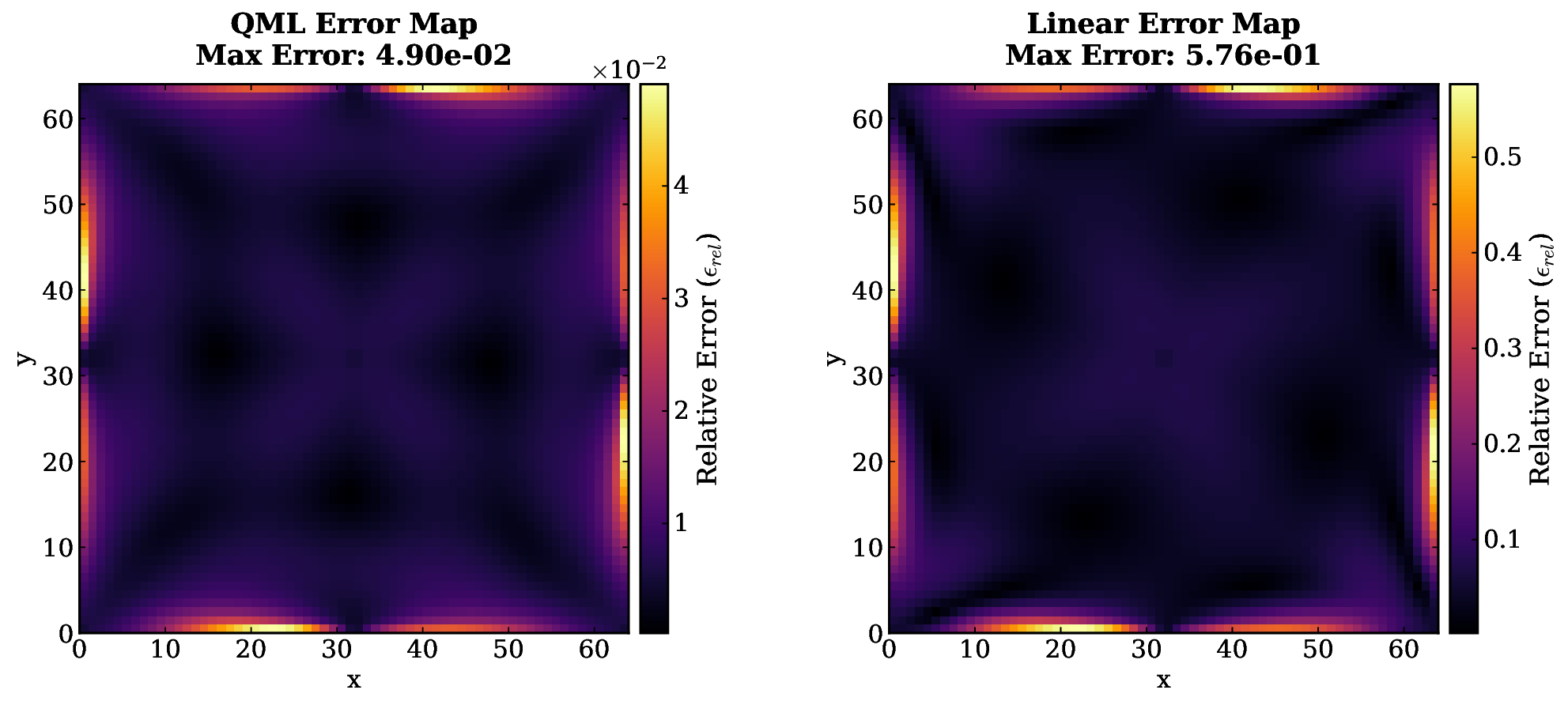}
        \caption{2D field of the velocity relative error for a simulation using the QML model learned (left) and the linear collision (right) compared with the analytical nonlinear LBM collision.}
    \end{subfigure}

    \caption{Taylor-Green vortex simulation using an analytical linear operator and a learned unitary for the nonlinear operator. The simulation parameters are: $64\times64$ lattice sites, $\tau=1$, $T=50$ time steps and $k_x=\frac{\pi}{N_X}$ using the quadratic velocity dependencies from $\eqref{eq:f_eq}$ compared to the same simulation up to linear order using the relative error for $u_{max}=0.1$. The loss function $L_{\rho_1}$ is used for the first register, while the cost function does not include the second register. The QML model has been trained for $4\cdot 10^6$ time steps, without observing barren plateaus (lower relative error can be obtained with longer training).}
    \label{fig:2r_TGV}
\end{figure*}
Figure~\ref{fig:2r_TGV} shows the high capability of the model when targeting the amplitudes of the first register. In contrast to previous cases for the unitary R1 model, we limited the training time steps to $4\cdot 10^6$ without encountering barren plateaus. This means the model can increase its accuracy roughly linearly over time during this phase of training. We also observe that the relative error map in b) matches the linear error pattern, suggesting that no trade-offs have been made and that the model is simply learning the nonlinear part of the collision operator with higher accuracy. To obtain these results (in contrast with the R1 nonunitary model), we have not lost the operator's unitary condition. Nonetheless, when we account for the phases using \eqref{eq:loss_rho_2}, the model's accuracy is drastically reduced. If we use a classical emulator that measures at each time step and set the final phase $\theta_f^t$ as the initial phase for the next time step, $\theta_i^{t+1}=\theta_f^t$, then the R2 model's accuracy is as high as the linear map for this test case. Only if we use \eqref{eq:loss_rho}  for the first register and reinitialise the phases of the second register at each time-step, we obtain a similar accuracy to Figure~\ref{fig:2r_TGV}.

\begin{figure}[htbp]
    \centering
    \includegraphics[width=0.5\textwidth]{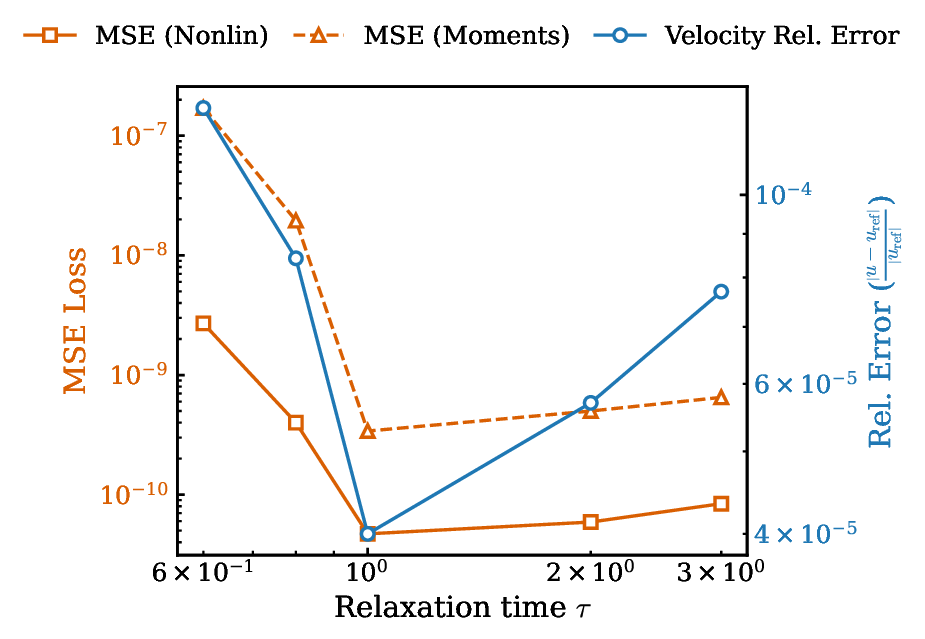}
    \caption{Comparison of the MSE of the predicted distribution function $f_i^{vqc}$ and post-linear collision distribution function $f_i^{lin}$ compared to $f_i^{ref}$ and momenta $[p_{xx}-p_{yy},p_{xy}]$ using $L=L_{A}$ as loss function. The figure also includes the absolute relative error of the velocity at the right axis. The simulation is carried at $u=0.05$.}
    \label{fig:2r_tau}
\end{figure}

The previous example, for a case with high velocity $u_{max}=0.1$, demonstrates the model's accuracy even at high velocities. Now, if we look at the dependency with the relaxation time $\tau$, we find a similar result as for the R1 model (See Figure~\ref{fig:tau_comparison} and Figure~\ref{fig:2r_tau}). As observed, the results do not change overall, despite having access to a copy of the initial distributions. The best results are obtained for $\tau=1$. As we previously introduced for the R1 model, for $\tau<1$, the MSE between the linear and nonlinear distribution functions is larger. Additionally, according to Figure~\ref{fig:spread_tau}, the operator becomes more nonunitary in this interval. Besides that, for $\tau>1$, non-conserving moments and out-of-equilibrium terms in the initial distribution $f^{lin}$ may make harder to conserve linear momentum for the VQC optimization procedure.

\section{Conclusions}
\label{sec:conclusions}
In this paper, we have thoroughly considered and tested variational quantum circuits to learn the nonlinear part of the collision operator in the Lattice Boltzmann Method (LBM). Previous research adopting a similar perspective mainly focuses on learning the collision step of LBM, restricted to unitary and nearly linear operators, and simulates a single time-step without emphasising the difference between linear and nonlinear contributions \cite{lactatus2025surrogate,Itani2025QMLLBM}. This work provides the first evidence that quantum circuits can effectively map post-collision linear distribution function $f_i^{lin}$ to predict the final distribution functions $f_i^{ref}$ at each lattice site after adding the $\mathcal{O}(u^2)$ terms.

To achieve this, we introduced two models. The first model, R1, which is the main focus of this article, uses an ansatz and encoding consistent with previous work \cite{lactatus2025surrogate} but specifically targets the complex mapping from the linear to the nonlinear collision operators. Our second model, R2, pioneers a dual-register architecture. This innovation grants the network exceptional flexibility, higher precision, and the ability to handle simulations with different relaxation times $\tau$, given its high decimal precision.

Our extensive testing of model R1 demonstrated the distinct advantage of predicting the nonlinear distribution function $f_i^{ref}$ from $f_i^{lin}$ is more effective than learning $f_i^{lin}$ directly from the pre-collision state $f_i^{str}$. Analysing the ratio $\eta$ (between the prediction MSE and the target MSE) alongside the non-unitarity of the operator at different velocities validated this approach. Evaluating the model across various velocities and viscosities revealed that R1 serves as a strong proof-of-concept, achieving peak accuracy across these conditions. Consistent with previous findings for training the full collision operator, the method is especially relevant for $\tau=1$, where the best results are observed. Regarding velocities $u$, higher velocities naturally increase the relative error in the final predicted velocities.

Furthermore, exploring the boundaries of R1 yielded crucial insights into the network's conservation mechanics. We observed an inverse relationship between the MSE of the quantum prediction $f_i^{vqc}$ and the relative error of the velocity $u$. We showed that increasing the weight $\lambda$ associated with the velocity's relative error in the cost function causes the predicted distribution to revert toward $f_i^{lin}$. This indicates that the model's capacity to conserve momentum at high accuracy is limited and cannot be boosted by changing the cost function without converging to $f^{lin}$. Illuminating how the VQC balances momentum conservation within a constrained topology. To expand these boundaries, we successfully demonstrated a non-unitary relaxation approach. By requiring the model to provide accurate amplitudes for only the first 9 elements of a 4-qubit statevector (neglecting the remaining 7), the model naturally decreases the total amplitude of these 9 elements from 1 to 0.5. This non-unitary approach provided highly accurate results across all tested Taylor-Green Vortex simulations. Internal analysis revealed that this success stems from the model shifting the difference between the linear and nonlinear distribution functions into the phases. While strictly constraining both amplitude and phase simultaneously poses challenges for $\tau \neq 1$, uncovering this phase-shifting behaviour is a significant step toward understanding how quantum circuits process nonlinearities.

Model R2, leveraging the dual-register architecture, demonstrated very high accuracy. For all tested Taylor-Green vortex simulations, R2 successfully predicted both amplitudes and phases. This success is contingent upon initializing the model with two uncorrelated copies of $f_i^{lin}$ and targeting the reduced density operators $\rho_1$ and $\rho_2$, such that the total density operator $\rho = \rho_1 \otimes \rho_2$ differs from the true total density operator of the uncorrelated target distribution functions $\rho_{ref}$. This approach proves that high-precision nonlinear modelling is highly attainable, effectively utilizing the entanglement of the initial product state and culminating in a natural measurement at each time-step to extract the highly accurate reduced density matrices. Future research will focus on studying the realisation of multiple time-steps within this accurate architecture.

Ultimately, this work establishes a foundational framework for predicting nonlinear distribution functions using variational quantum circuits. By demonstrating that quantum algorithms can successfully navigate the complexities of nonlinear fluid dynamics, we have opened a highly promising pathway toward industrially relevant, multi-step quantum simulations. Building directly on these findings, future research will focus on extending the multi-step capabilities of both the non-unitary R1 and the dual register R2 models. Furthermore, identifying specific fluid dynamics regimes where the unitary R1 architecture exhibits a definitive quantum advantage over classical methods remains an exciting frontier. Expanding this quantum machine learning paradigm to the Carleman collision operator, which offers a natively linear yet non-unitary structure, presents another compelling avenue for investigation. Finally, integrating advanced data encodings inspired by recent literature \cite{Lehmann2022} will be instrumental in maximizing the precision of single-register models, driving this innovative approach ever closer to large-scale, practical application.

\section*{Acknowledgment}
V.L. acknowledges the support from the Research Council of Finland (Flagship of Advanced Mathematics for Sensing Imaging and Modelling grant 359183).

\FloatBarrier
\appendix
\section{Diffusive scaling}
\label{sec:appendix}
In this appendix, we introduce the diffusive scaling as an efficient alternative to using larger lattice velocities in the context of QML for LBM, thereby benefiting from lower error rates in predicting the nonlinear term at lower velocities.

Diffusive scaling in LBM involves a trade-off among the number of time steps, the number of lattice sites, and the velocity. We define the physical time-step $\Delta t$ and the physical lattice-step $\Delta x$ using the viscosity as a constant parameter: $\nu = \frac {\Delta x^2}{\Delta t} = const$. Given that the Mach number $Ma=\frac{u}{c_s}$ and the sound velocity in lattice units is $c_s^{lat}=O(1)$, then we have
\begin{equation}
    u^{lat}=\frac{u^{phys}\Delta t}{\Delta x}=u^{phys} \Delta x
\end{equation}
This means that when $\Delta x\to 0$ then $Ma\to 0$, which guarantees convergence, as LBM error is typically of the order of $O(Ma^3)$. To calculate how the computational complexity changes when setting the viscosity as constant and modifying $\Delta x$ and $\Delta t$, we will consider the scaling factor $\beta$. We will call the original parameters before implementing the diffusive scaling as the base variables and the new converted parameters as the scaled variables. Therefore considering $\Delta x_S=\beta \Delta x_B$ and $\Delta t_S=\beta^2 \Delta x_B$. Then, the physical variables of the scaled basis become
\begin{equation}
\begin{aligned}
c_s^{phy}{_S} &= \frac{\Delta x_S}{\Delta t_S} 
      = \frac{1}{\beta} \frac{\Delta x_B}{\Delta t_B} 
      = \frac{1}{\beta} \, c_s^{phy}{_B}, \\[2mm]
u^{lat}_S &= u^{phy}\frac{\Delta t_S}{\Delta x_S} 
      = u^{phy}\beta \frac{\Delta t_B}{\Delta x_B} 
      = \beta \, u^{lat}_B, \\[2mm]
P_S &= \frac{\Delta x_S^2}{\Delta t_S^2} 
      = \frac{1}{\beta^2} \frac{\Delta x_B^2}{\Delta t_B^2} 
      = \frac{1}{\beta^2} \, P_B, \\[1mm]
\nu_S & \coloneqq \nu_B.
\end{aligned}
\end{equation}
Notice that the lattice sound velocity $c_s^{lat}$ is constant for LBM. In the same way, the physical velocity $u^{lat}$ to be simulated is a constant imposed in the simulation. This means that the $Ma$ is reduced in the scaling, as seen previously. Another consideration is the algorithm's computational complexity. When scaling the computational variables with $\Delta x\to 0$, the number of time-steps and number of lattice sites scale as well to
\begin{equation}
\begin{aligned}
N_S &= \frac{L^D}{ \Delta x_S^D} 
      = \frac{1}{\beta^D}\frac{L^D}{\Delta x_B^D}
      = \frac{1}{\beta^D} \, N_B, \\[2mm]
T_S &= \frac{t}{\Delta t_B} 
      = \frac{1}{\beta^2}\frac{t}{\Delta x_S^2}
      = \frac{1}{\beta^2} \,T_B , \\[2mm]
\end{aligned}
\end{equation}
with $D$ the number of dimensions. As $0<\beta\leq1$ for the diffusive scaling, the number of lattice sites $N_S$ and the number of time-steps $T_S$ increase the computational cost of the simulation. As seen in Table~\ref{tabl:complexity}, for $D=2$, the computational complexity in the quantum computer is higher than in the classical counterpart. 
\begin{table}[htbp]
\centering
\begin{tabular}{|c|c|c|}
\hline
 & Classical & Quantum \\ 
\hline
Base 
& $O(T_B N_B)$ 
& $O\!\left(\log^2(N_B)\, T_B\right)$ \\ 
\hline
Scaled 
& $O\!\left(\dfrac{T_B N_B}{\beta^4}\right)$ 
& $O\!\left(\log^2\!\left(\dfrac{N_B}{\beta^2}\right)\, \dfrac{T_B}{\beta^2}\right)$ \\ 
\hline
\end{tabular}
\caption{Computational complexity comparison between classical and quantum LBM implementations for the base and scaled algorithms in $D=2$.}
\label{tabl:complexity}
\end{table}

\begin{figure}
    \centering
        \includegraphics[width=0.48\textwidth]{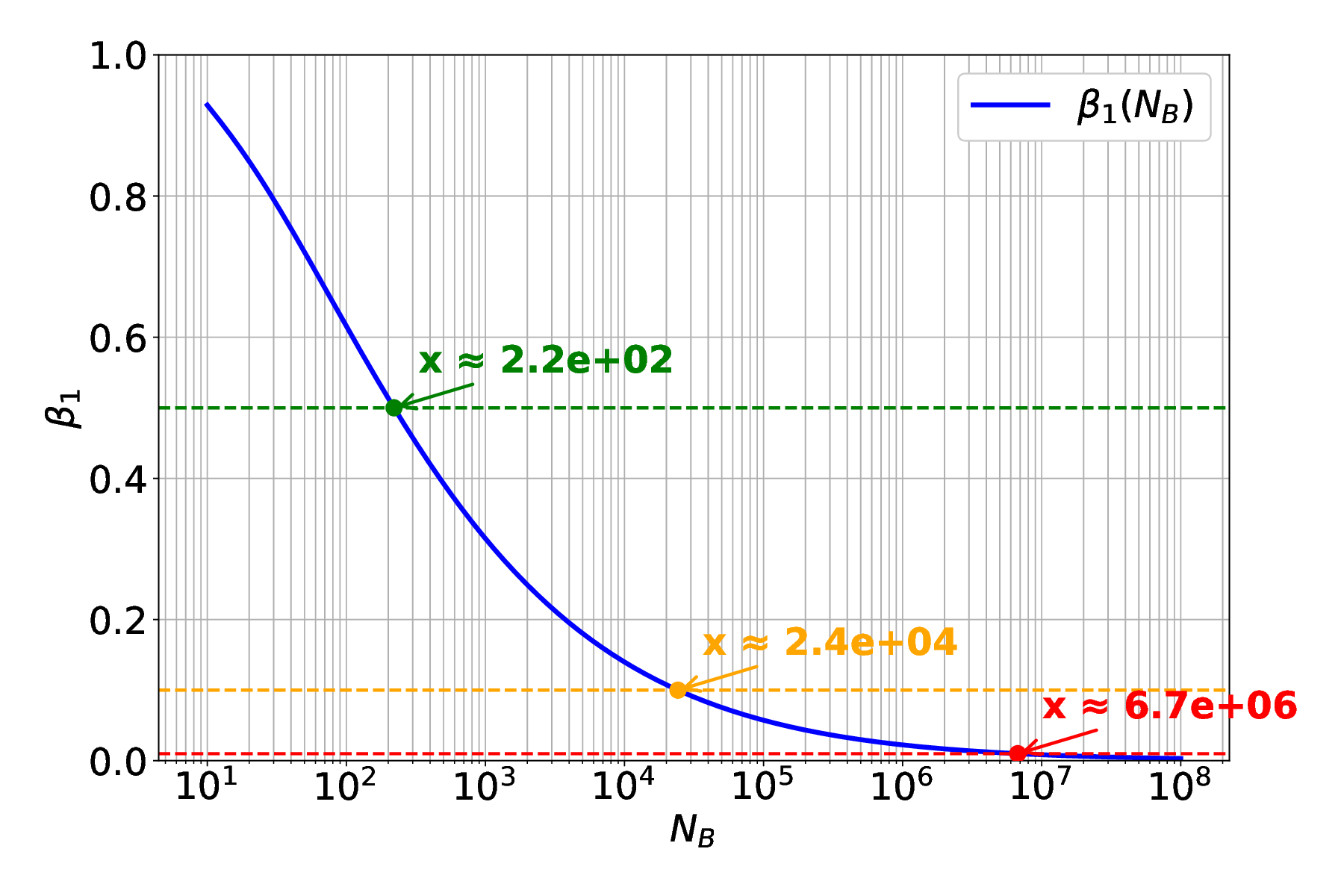}
    \caption{Graphical representation of the function $\beta_1(N_B)$. Three lines has been drawn in the map from top to bottom: The first line (green) with $\beta_1=0.5$, the second (yellow) with $\beta_1=0.1$ and the third (red) with $\beta_1=0.01$.}
    \label{fig:beta1}
\end{figure}

The next step is to calculate the ratio between the base-classical case with the quantum-scaled case to know if the scaled case is advantageous 
\begin{equation}
\eta=\frac{T_B N_B}{\log^2\Big(\dfrac{N_B}{\beta^2}\Big) \dfrac{T_B}{\beta^2}}=\frac{N_B \beta^2}{log^2(\frac{N_B}{\beta^2})}
\end{equation}
If we find the expression of $\beta(N_B)$ at which $\eta=1$, then we can know which regime provides a quantum advantage. To do that we will consider 
\begin{equation}
   \beta \sqrt{N_B}=log(\frac{N_B}{\beta^2})=log(N_B)-2log({\beta)}
\end{equation}
\begin{equation}
    \beta_0\approx \frac{log(N_B)}{\sqrt{N_B}}
\end{equation}
Now, if we consider
\begin{equation}
       \beta_1 \sqrt{N_B}=log(N_B)-2log({\beta_0)}
\end{equation}
we obtain
\begin{equation}
    \beta_1= \frac{2(log(N_B)-log(log(N_B))}{\sqrt{N_B}}
\end{equation}
which set the lower limit at which the diffusive scaling used in the quantum algorithm loses its advantage. As we can see in Fig~\ref{fig:beta1}, when $N_B>2.4 \cdot 10^{4}$, we can reduce the lattice velocity $u^{lat}$ ten times with similar scaling as the classical algorithm, corresponding to only $160\times160$ lattice sites in 2D. For $N_B>6.7 \cdot 10^{6}$ ($2600\times2600$), the lattice velocity can be reduced up to 100 times, conserving advantage. While the figure takes into account $\beta_1$ for $D=2$, a similar result can be obtained for $D=3$

Two different regimes are then available for the quantum machine learning approach. First, the Reynols scaling, where we set a lattice velocity $u^{lat}<0.1$ and increase the number of lattice sites. As $\Delta x$ is constant, the Reynolds number $Re=\frac{Lu}{\nu}$ increases. We can obtain high Reynolds number simulations, but we are limited in the maximum velocity to simulate, and the leading error rate $O(Ma^3)$ can be industrially limiting, especially at the higher band. In contrast, the diffusive scaling, while more computationally expensive (with a higher number of sites $N$ and time-steps $T$ not translating to higher $Re$), yields higher physical velocities to be simulated, and the error rate caused by LBM is negligible. In this incompressible regime, especially advantageous for a very large number of lattice sites, industrially relevant applications are possible, such as wind turbines, bullet trains, skyscrapers, pipelines, cruise ships, landing gear noise, and others. However, the error rate generated by the variational quantum circuit dominates over the leading error rate $O(Ma^3)$ \cite{Itani2025QMLLBM}. Therefore, the use of this strategy should be considered as a tool for cases where the total error, given by the error rate at low velocities and more time-steps, is lower than that for larger velocities, which produce larger errors per time-step but require fewer time-steps during the simulation. 
\bibliography{apssamp}

\end{document}